\documentstyle[12pt]{article}
\begin{document}
\thispagestyle{empty}
\begin{center}
{\large\bf Negative-Energy Perturbations \\
 in Circularly Cylindrical Equilibria \vspace{1mm}\\
 within the Framework of Maxwell-Drift Kinetic Theory} \vspace{6mm}\\
{\large G. N. Throumoulopoulos$^{\dag}$
\ \ and D. Pfirsch$^{\star}$\\ 
\vspace{3mm}
{\sl $^{\bf {\dag}}$ Division of Theoretical Physics\\
 Department of Physics, University of Ioannina\\ 
P. O. Box 1186, GR  451 10 Ioannina, Greece \\ \vspace{2mm}
  ${\bf ^{\star}}$ Max-Planck-Institut f\"{u}r Plasmaphysik, EURATOM
Association \\\vspace{1mm}
 D-85748 Garching, Germany}  }
\end{center}
\newpage
\thispagestyle{empty}
\begin{center}
{\large\bf Abstract} 
\end{center}

The conditions for the existence of negative-energy perturbations (which could
be nonlinearly unstable and cause anomalous transport)
are investigated  in the framework of linearized 
collisionless Maxwell-drift kinetic theory
for the case of  equilibria of 
magnetically confined, circularly cylindrical plasmas 
and vanishing initial field
perturbations.  
  For wave vectors with a non-vanishing
component parallel to the magnetic field, 
the plane equilibrium conditions
(derived by Throumoulopoulos and Pfirsch  
[Phys Rev. E {\bf 49}, 3290 (1994)]) 
are shown to
remain valid, while the condition for 
  perpendicular perturbations  (which are found to be the most important modes)
is   modified.
Consequently, besides the tokamak equilibrium regime in which  
the existence
of negative-energy perturbations is related to the 
threshold value of 2/3 of the quantity
$\eta_\nu = 
\frac {\textstyle  \partial \ln T_\nu} {\textstyle  \partial \ln N_\nu}$,
 a new regime appears, not present in plane equilibria, in which
negative-energy perturbations exist for {\em any} value of $\eta_\nu$. 
 For various  analytic cold-ion tokamak equilibria
 a substantial fraction of thermal electrons
  are associated with negative-energy perturbations (active particles).
 In particular,
for linearly stable equilibria of a paramagnetic plasma with flat  electron
temperature profile ($\eta_e=0$), the entire  velocity space 
is occupied by active electrons.
The part of the velocity space occupied 
 by active particles increases
from the center to the plasma edge  and is larger in a paramagnetic plasma than 
in a diamagnetic plasma with the same pressure profile.
It is also shown that, unlike in plane
equilibria, negative-energy perturbations exist in force-free reversed-field
pinch equilibria with a substantial fraction of active particles.
	The present results, in particular   the fact that a threshold value of 
$\eta_\nu$ is not necessary for the existence of negative-energy perturbations,
 enhance even more the relevance of these modes.
\newpage
\setcounter{page}{1}
\begin{center} 
{\bf I.\ \ INTRODUCTION }
\end{center}
Negative-energy perturbations are potentially dangerous because they may become nonlinearly
unstable and cause anomalous transport \cite{We} - \cite {PfWe}. 
Conditions for the existence of perturbations of this kind 
can be obtained on the
basis of the expressions for the second variation of the free energy which were
derived by Pfirsch and Morrison \cite{PfMo} for arbitrary perturbations of
general equilibria within the framework of collisionless Maxwell-Vlasov and
Maxwell-drift kinetic theories.

For  homogeneous, magnetized plasmas and  vanishing initial
field perturbations they found that negative-energy perturbations exist for
any wave vector ${\bf k}$ having a non-vanishing component parallel to
the magnetic field (parallel and oblique modes) whenever the condition
\begin{equation} 
v_{\parallel} \frac{\partial f^{(0)}_{g\nu}}{\partial v_{\parallel}}
>0                                                        \label{eq:eg1}
\end{equation} 
holds for the equilibrium guiding center distribution function $f^{(0)}_{g\nu}$
for some particle species $\nu$ and parallel velocity $v_{\parallel}$ in the
frame of lowest equilibrium energy. 
For  inhomogeneous magnetically confined
plasmas with equilibria depending on just one Cartesian coordinate $y$,
Throumoulopoulos and Pfirsch \cite{ThPf} showed that, 
in addition to parallel and
oblique modes, for which condition (1)  applies, perpendicular modes have
also negative energies if 
\begin{equation} 
\frac{dP^{(0)}}{dy}   \frac{\partial  f^{(0)}_{g\nu}}{\partial y}
<0,                                                    \label{2}
\end{equation} 
holds,
where $P^{(0)}$ is the equilibrium plasma pressure.  For tokamaklike
equilibria, condition (2) implies a threshold value of 2/3 of the quantity
$\eta_{\nu} = 
\frac{\textstyle  \partial \ln T_{\nu}}{\textstyle  \partial \ln N_{\nu}}$,
where $T_{\nu}$ is the temperature and $N_{\nu}$ the density of  particle
species $\nu$.  These investigations are extended in this paper to the more
interesting case of  circularly cylindrical plasmas. 
The method of investigation
consists in evaluating the general expression for the second-order perturbation
energy obtained by Pfirsch and Morrison 
 within the framework of the linearized collisionless Maxwell-drift kinetic
theory. The most important conclusions are: 
\begin{enumerate}
\item  Condition (1) for the existence
of parallel and oblique modes remains valid.
\item For tokamak and reversed-field pinch cold-ion equilibria a new regime
appears, not present in plane equilibria, 
in which perpendicular negative-energy
perturbations exist without restriction on the values  of $\eta_\nu$.
\end{enumerate}

The equilibrium properties of the circularly cylindrical plasmas under
consideration are discussed in Sec. II. The second-order 
perturbation energy for
vanishing initial field perturbations is presented in Sec. III. The
relevant lengthy derivation is reported in Appendix A. 
The conditions for the
existence of negative-energy perturbations are obtained in Sec. IV. 
The cases of
parallel, oblique and perpendicular wave propagation 
are examined separately. The
consequences of the condition for the existence of
 perpendicular negative-energy
perturbations in straight tokamak and reversed-field pinch equilibria 
are discussed in Sec. V.
For various analytic cold-ion equilibria with non-negative 
and negative values of $\eta_e$, 
the part of the velocity space occupied by electrons associated with
negative-energy perturbations is also obtained. Two examples are presented in
Appendix B.
 The main results are summarized in Sec. VI.
\begin{center}
{\bf II.\ \ EQUILIBRIUM}
\end{center}

The collisionless Maxwell-drift kinetic theory 
 applied in the present paper is based on  Littlejohn's Lagrangian
formulation of the guiding center theory   \cite{Li} 
in the form given by Wimmel \cite{Wim}.
A brief review of this theory is  given in  the
first paragraph of Sec. III. More details can be found in Ref. \cite{PfMo}
and in Sec. II of Ref. \cite{ThPf}.

For a magnetically confined, circularly cylindrical plasma
the equilibrium  vector potential and  magnetic field
are given by 
\begin{equation}  
{\bf A}^{(0)} =A^{{(0)} }_{\theta}(r){\bf e}_{\theta}+A^{{(0)} }_{z}(r)
{\bf e}_{z}                                                     
                                                      \label{36}
\end{equation} 
and
\begin{equation} 
{\bf B}^{{(0)} } = B^{{(0)} }_{\theta}(r)
{\bf e}_{\theta} + A^{{(0)} }_{z}(r)
{\bf e}_{z},                                                    
                                  \label{37}
\end{equation} 
with
\begin{equation}  
\frac{1}{r}\frac{d}{dr} (rA^{{(0)} }_{\theta}) = B^{{(0)} }_{z}, 
\ \ \  (A^{{(0)} }_{z})^{\prime} = - B^{{(0)} }_{\theta}.
                                     \label{38}
\end{equation} 
Here, $r$, $\theta$, $z$ are cylindrical coordinates with unit base vectors
 ${\bf e}_r$, ${\bf e}_\theta$, ${\bf e}_z$ and 
 the
prime denotes differentiation with respect to $r$.  It is assumed
that there is no equilibrium electric field. To calculate the guiding center
velocity, Eq. (\ref{R 24}) below,  one needs the following quantities: 
\begin{equation} 
{\bf b}^{{(0)} } =\frac{{\bf B}^{(0)} }{B^{{(0)} }}
=\frac{B^{{(0)} }_{\theta}}{B^{{(0)} }}{\bf e}_{\theta}+
\frac{B^{{(0)} }_{z}}{B^{{(0)} }}{\bf e}_{z} =
b^{{(0)} }_{\theta} {\bf e}_{\theta} +
b^{{(0)} }_{z}{\bf
e}_z,                                              
                                                  \label{39}
\end{equation} 
\begin{equation}  
{\bf A}^{\star{(0)} }_{\nu} = {\bf A}^{{(0)} } +
 \frac{m_{\nu}c}{e_\nu}v_\parallel 
 {\bf b}^{{(0)} },
                                                    \label{40}
\end{equation} 
\begin{equation}  
e_{\nu} \phi^{\star{(0)} }_{\nu} = \mu B^{{(0)} } + 
\left(\frac{m_{\nu}}{2}\right)
v_\parallel^2 
                                                 \label{41},
\end{equation} 
 \begin{equation}  
{\bf  v}^{{(0)} }_{E} = c \frac{{\bf E}^{{(0)} }
\times{\bf B}^{{(0)} }}{\left(B^{{(0)} }\right)^2} 
={\bf 0},
                                      \label{42}
\end{equation} 
\begin{equation}  
{\bf E}^{\star{(0)} }_\nu = 
- {\bf \nabla} \phi^{\star{(0)} }_{\nu}= 
- \frac{\mu}{e_{\nu}} \left(B^{{(0)} }\right)^{\prime}{\bf e}_r 
                                               \label{43}
\end{equation} 
and
\begin{equation} 
{\bf B}^{\star{(0)} }_{\nu} = {\bf\nabla}\times
{\bf A}^{\star{(0)} }_{\nu} = B^{\star{(0)} }_{\nu\parallel} {\bf b} +
\frac{m_{\nu}c}{e_\nu} v_\parallel\frac{\left(b^{(0)} _\theta\right)^2}{r}
 \left({\bf e}_r\times{\bf b}^{(0)} \right),
                                       \label{44}
\end{equation} 
with
\begin{equation} 
B^{\star{(0)} }_{\nu\parallel} = {\bf B}^{\star{(0)} }_{\nu} 
\cdot {\bf b}^{{(0)} } =
B^{{(0)} } + \frac{m_\nu c}{e_\nu} v_\parallel Y_{\theta z} 
                                        \label{45}
\end{equation} 
and
\begin{equation} 
 Y_{\theta z}(r)\equiv {\bf b}^{{(0)} } \cdot \left({\bf\nabla} \times
{\bf b}^{{(0)} }\right) = \left( b^{{(0)} }_{\theta}\right)^\prime 
 b^{(0)} _z - \left(b^{{(0)} }_z\right)^{\prime} b^{{(0)} }_{\theta}
+ \frac{b_{\theta}^{(0)}  b_z^{(0)} }{r}.
\label{46}
\end{equation} 
With the aid of Eqs. (\ref{39}-\ref{46}) 
the guiding center velocity  takes the
form 
\begin{equation}  
{\bf  v}^{{(0)} }_{g\nu} = v_\parallel {\bf b}^{{(0)} } -
\frac{\mu c}{e_{\nu}B^{\star{(0)} }_{\nu\parallel}} 
\frac{d B^{(0)} }{d r}
\left({\bf e}_r \times {\bf b}^{{(0)} }\right) + 
\frac{v_\parallel^2 }{\omega^\star_\nu} 
\frac{\left(b^{(0)}_\theta\right)^2}{r}
\left({\bf e}_r \times {\bf b}^{{(0)} }\right),
                                                    \label{47}
\end{equation} 
with
$
\omega^{\star}_{\nu} \equiv
\frac{\textstyle   e_\nu B^{\star{(0)} }_{\nu\parallel}}
{\textstyle   cm_{\nu}}.
$
The first, second and third terms in
(\ref{47}) are the component of ${\bf  v}^{{(0)} }_{g\nu}$ parallel to ${\bf
B}^{{(0)} }$, the grad-B drift and the curvature drift.  
${\bf  v}^{{(0)} }_{g\nu}$
has no $r$-component and therefore $r$ is a constant of motion.  Since there
is also no force parallel to ${\bf B}^{{(0)} }$, another constant of
motion is the parallel guiding center velocity $ v_\parallel$. 
 The guiding center
distribution functions $f^{{(0)} }_{g\nu}$ are therefore  functions of $r$, $
v_\parallel$ and the magnetic moment $\mu$.

To calculate the current density ${\bf  J}^{{(0)} }$, we apply the
general formula (8.15) of Ref. \cite{CoPfWi}, which was derived
in the context of collisionless Maxwell-drift kinetic theory.  The result is
\begin{eqnarray} 
{\bf  J}^{{(0)} }& = &\frac{c}{4\pi} {\bf\nabla} \times 
{\bf B}^{{(0)} }                  \nonumber \\ 
&=& \sum_{\nu} e_\nu\int d v_\parallel d\mu 
B^{\star{(0)} }_{\nu\parallel} f^{{(0)} }_{g\nu} {\bf  v}_{g\nu}\nonumber \\ 
 &&-\sum_{\nu} c{\bf  \nabla}\times 
\int d v_\parallel d\mu\left\{
B^{\star{(0)} }_{\nu\parallel}f^{{(0)} }_{g\nu} \left(\mu{\bf b} -
\frac{m_\nu}{B} v_\parallel \bf  v_{g\nu\bot}\right)\right\},
                                      \label{48}
\end{eqnarray} 
where ${\bf  v}_{g\nu\bot} = {\bf  v}_{g\nu} -
v_\parallel {\bf b}$.  The first and 
second sums in (\ref{48}) represent, respectively, the guiding center and the
magnetization contributions to ${\bf J}^{(0)} $.  
Taking the cross product of Eq.  (\ref{48}) with 
${\bf B}^{(0)}$ and using Ampere's low on the left-hand side of the
resulting equation we obtain after  some straightforward algebraic
manipulations 
 \begin{equation} 
\frac{d}{dr} \left[P^{{(0)} } + \frac{B^{{(0)} }}{8{\pi}}\right] +
\frac{(B^{{(0)} }_{\theta})^2}{4{\pi}r}
+{\Pi}(r)=0,                                            
                                                    \label{51}
\end{equation} 
with
\begin{equation} 
P^{{(0)} } = \sum_{\nu} \int d  v_\parallel d\mu \ 
{\mu} B^{{(0)} } B^{\star{(0)} }_{\nu\parallel} f^{{(0)} }_{g\nu}
                                           \label{52}
\end{equation} 
and
\begin{eqnarray} 
{\Pi}(r) &\equiv& \sum_{\nu} \int d  v_\parallel 
d\mu B^{\star{(0)} }_{\nu\parallel}
\frac{\left(b^{(0)}_\theta\right)^2}{r} 
\left( {\mu}B^{(0)} -m_{\nu} v_\parallel^2 \right)
f^{{(0)} }_{g\nu} \nonumber \\ 
&&-2\sum_{\nu} \frac{m_\nu c}{e_{\nu}}\int d v_\parallel d\mu v_\parallel
\left\{\    
 \frac{b^{{(0)} }_{\theta}b^{{(0)} }_z}{r}
\left({\mu} \left(B^{(0)} \right)^{\prime}\right.\right.\nonumber \\ 
 & & \left.\left. -m_\nu  v_\parallel^2   
\frac{\left(b^{(0)}_\theta\right)^2}{r}\right)\right\}.
                                            \label{53}
\end{eqnarray} 
Relation (\ref{51}) can also be derived by the momentum-conservation
relation $T^{\mu}_{\rho\ , \mu}=0$ with the tensor
$T^{\mu}_{\rho}$ given in explicit form by Eq. (76) of
Ref. \cite{PfMo85}.  (The comma in the subscript denotes covariant
derivative.)  For Maxwellian distribution functions it holds that
${\Pi}_{\nu}=0$, and Eq. (\ref{51}) reduces to the known MHD
equilibrium relation. 
\begin{center}  
{\bf III.\ \ SECOND-ORDER PERTURBATION ENERGY}
\end{center}
 The second-order energy of perturbations
around an equilibrium state is given by
\begin{equation}
F^{(2)} = \int d^3 x \ T_0^{(2)0},           \label{P0}
\end{equation}
where  $T_0^{(2)0}$ is the energy component of the second-order
energy-momentum \newline tensor \cite{PfMo}
\begin{eqnarray}
T_\rho^{(2)\lambda}&=& -\sum_\nu\int d\hat{\tilde{q}} d\tilde{P}\left(
    \frac{\partial S_\nu^{(1)}}{\partial\tilde{q}^\rho}-\frac{e_\nu}{c}A_\rho
    ^{(1)}\right)\left[ f_\nu^{(0)}\left(\frac{\partial S_\nu^{(1)}}
    {\partial\tilde{q}^\kappa}-\frac{e_\nu}{c}A_\kappa^{(1)}\right)
    \frac{\partial^2{\cal H}_\nu^{(0)}}
    {\partial \tilde{P}_\lambda\partial\tilde{P}_\kappa}
    \right.
    \nonumber \\ 
& & \left.+f_\nu^{(0)}F_{\tau\sigma}^{(1)}
     \frac{\partial^2{\cal H}_\nu^{(0)}}
    {\partial\tilde{P}_\lambda\partial F_{\tau
     \sigma}^{(0)}}
     +\left(f_\nu^{(0)}
     \frac{\partial S_\nu^{(1)}}{\partial \tilde{P}_i}\right)_{\ ,i}
     \frac{\partial {\cal H}_\nu^{(0)}}
          {\partial\tilde{P}_\lambda}\right] \nonumber \\ 
& &  -2F_{\mu\rho}^{(1)}\sum_\nu\int d\hat{\tilde{q}} d\tilde{P}\left[
     f_\nu^{(0)}\left(\frac{\partial S_\nu^{(1)}}{\partial\tilde{q}^\kappa}
    -\frac{\textstyle e_\nu}{\textstyle  c} A_\kappa^{(1)}\right)
     \frac{\partial^2{\cal H}_\nu^{(0)}}
     {\partial\tilde{P}_\kappa\partial F^{(0)}
     _{\mu\lambda}}\right. \nonumber \\ 
& &  \left.+f_\nu^{(0)} F_{\sigma\tau}^{(1)}
    \frac{\partial^2{\cal H}_\nu^{(0)}}
    {\partial F_{\mu\lambda}^{(0)}\partial F_{\sigma\tau}
    ^{(0)}}\right]
    -\frac{1}{4\pi} F_{\mu\rho}^{(1)} F^{(1)\mu\lambda}
    \nonumber \\
& &
     +\delta_\rho^\lambda
   \left(\sum_\nu\int d\hat{\tilde{q}} d 
   \tilde{P} f_\nu^{(0)}({\cal H}_\nu^{(2)}
   -{\cal H}_\nu^{(0)(2)}) + \frac{1}{16\pi} F_{\tau\sigma}^{(1)}
    F^{(1)\tau\sigma}\right).
                                                      \label{P1}
\end{eqnarray}
 Here, the superscripts $(0)$, $(1)$ and $(2)$, respectively, denote 
equilibrium first-and second-order quantities; $A_\rho=\left(-\phi,\
{\bf A}\right)$, where $\phi$ is the scalar potential and ${\bf A}$ 
the vector potential of the electromagnetic field; $F_{\mu\nu}$ is the
electromagnetic tensor; $S_\nu^{(1)}$ are generating functions
associated with the perturbations; 
the scalar quantity $\left(f_\nu^{(0)}
     \frac{\textstyle  \partial S_\nu^{(1)}}{\textstyle 
          \partial \tilde{P}_i}\right)_{\ ,i}$
 results from the  contraction in the  second-order tensor
$\left(f_\nu^{(0)}
     \frac{\textstyle  \partial S_\nu^{(1)}}
          {\textstyle  \partial \tilde{P}_i}\right)_{\ ,j}$;
the rest of the notation is defined on page 273 of 
Ref. \cite{PfMo}.\newline
 In expression
(\ref{P1}) the time derivatives $\frac{\textstyle  
 \partial S_\nu^{(1)}}{\textstyle  
\partial t}$ are  given by 
\begin {equation} \frac{\partial
S_\nu^{(1)}}{\partial t}-e_\nu A_0^{(1)} =-\lbrack S_\nu^{(1)},
  H_\nu^{(0)}\rbrack 
  +\frac{\textstyle e_\nu}{\textstyle  c}
        {\bf A}^{(1)}\cdot\frac{\partial H_\nu^{(0)}}{\partial{\bf P}}
    -F_{\mu\lambda}^{(1)}\frac{\partial H_\nu^{(0)}}
       {\partial F_{\mu\lambda}^{(0)}},
                                                        \label{P2}
\end{equation}
where the mixed variable Poisson bracket is defined as
$$ \lbrack a,b\rbrack=\frac{\partial a}{\partial \tilde{q}_i}
    \frac{\partial b}{\partial \tilde{P}_i} - 
     \frac{\partial a}{\partial \tilde{P}_i}
    \frac{\partial b}{\partial \tilde{q}_i}.$$
The Hamiltonian for the guiding center motion of particle species $\nu$ is
obtained from the Lagrangian 
\begin{equation} 
L_\nu=\left(\frac{e_\nu}{c}\right){\bf A}_\nu^{\star}
\cdot\dot{{\bf x}} - e_\nu\phi_\nu^\star 
                                                \label{P3}
\end{equation} 
with
\begin{displaymath} 
{\bf A}^{\star}_{\nu} = {\bf A} + \frac{m_{\nu}c}{e_\nu}q^4
 {\bf b},
\end{displaymath} 
\begin{displaymath} 
e_{\nu} \phi^{\star}_{\nu} = e_\nu\phi + \mu B +
\frac{m_{\nu}}{2} \left((q^4)^2 +{\bf v}_E^2 \right),
\end{displaymath} 
\begin{displaymath} 
{\bf  v}_{E} = c \frac{{\bf E}
\times{\bf B}}{B^2}, 
\end{displaymath} 
\begin{displaymath} 
{\bf E}=-{\bf \nabla}\phi - \frac{1}{c}\frac{\partial{\bf A}}{\partial t},
\ \ \ \ {\bf B }= \nabla\times {\bf A},\ \  \ \ {\bf b} = \frac{{\bf B}}{B}.
\end{displaymath} 
This Lagrangian  is defined in terms of the variables
$$ t,\ \ {\bf x}={\bf x}\left(q^1,q^2,q^3\right)
\ \ \mbox{and}\ \ q^4. $$
Here, $q^1,q^2,q^3$ are generalized coordinates in normal space and $q^4$
is an additional independent variable for wich one of the Lagrangian equations
yields the relation $q^4={\bf v}\cdot{\bf b}=v_\parallel$. 
 The momenta canonically conjugated
to ${\bf x}$ and $q^4$ follow from (\ref{P3}) as
\begin{equation}
{\bf p}=\frac{\partial L_\nu}{\partial \dot{{\bf x}}}
    =\frac{\partial L_\nu}{\partial\dot{q}^l}{\bf e}^l
    =\frac{e_\nu}{c}{\bf A}^{\star}_\nu,   \ \  \ 
p_4 = \frac{\partial L_\nu}{\partial\dot{q}^4}= 0,
                                                       \label{P4}
\end{equation}
where 
 ${\bf e}^l$ 
are the reciprocal base vectors. Since Eqs. (\ref{P4}) do not contain
$\dot{{\bf x}}$ and  $\dot{q}^4$, they are constraints between the momenta 
and the coordinates. It therefore follows  that
Hamilton's equations based on the usual  Hamiltonian corresponding to
the above non-standard Lagrangian
are not the equations of motion. To overcome this difficulty,
Dirac's  theory of constrained dynamics
\cite{Di}
 is applied,  which yields the Dirac Hamiltonians:
\begin{equation}
 H_\nu=e_\nu\phi_\nu^\star + {\bf v_{g\nu}}\cdot
       \left({\bf p}-(e_\nu/c){\bf A}^\star_\nu\right)+V^4p_4,
                                                            \label{P5}
\end{equation}
from which
\begin{equation}  
\dot{{\bf x}}={\bf v}=\frac{\partial H_\nu}{\partial {\bf p}}=
              {\bf v}_{g\nu}\left(t,{\bf x},q^4\right) 
             = \frac{q^4}{ B^\star_{\nu\parallel}}
              {\bf B}_\nu^\star + \frac{c}{B^{\star}_{\nu\parallel}}
              {\bf E}^\star_\nu\times{\bf b}
                                             \label{R 24}
\end{equation}  
and
\begin{equation} 
 \dot{q}^4 = \frac{\partial H_\nu}{\partial p_4}=V^4 
                 \left(t,{\bf x},q^4\right) =
           \frac{e_\nu}{m_\nu}
           \frac{1}{B^\star_{\nu\parallel}}
           {\bf E}_\nu^\star\cdot
           {\bf B_\nu}^\star
                                             \label{R 25}
\end{equation} 
follow.
(Here, ${\bf E}^\star_\nu \equiv {\bf \nabla} \phi_\nu^\star - 
\frac{\textstyle   1}{\textstyle   c}
        \frac{\textstyle  \partial{\bf A}^\star_\nu}
              {\textstyle  \partial t},  \ \ \ 
{\bf B_\nu}^\star\equiv {\bf \nabla} \times{\bf A}^\star_\nu  \ \ 
         \mbox{and} \ \ 
B^{\star}_{\nu\parallel} = {\bf B}^\star_{\nu}\cdot {\bf b}$.)
Special solutions of the equations of motion following
from the Hamiltonians (\ref{P5}) are  the
constraints (\ref{P4}). The distribution functions
    $f_\nu({\bf x}, q^4, {\bf p},
p_4, t)$ must guarantee that these constraints are satisfied. As concerns this
requirement, 
it is important to note that ${\bf p}-
      \left(\frac{\textstyle   e_\nu}{\textstyle  
c}\right){\bf A}^\star_\nu=0$ and $p_4=0$ do not represent special values
 of some
constants of motion. Therefore, $\delta$-functions of the constraints are not
constants of motion either. On the other hand,  $f_\nu$ must
be proportional to
such $\delta$-functions and, at the same time,
  also  a constant of motion. Both
conditions are uniquely satisfied by
\begin{equation}
f_\nu=\delta(p_4)\delta
      \left({\bf p}-\frac{e_\nu}{c}{\bf A}^\star_\nu\right)
       B_{\nu\parallel}^\star
       f_{g\nu}\left({\bf x},q^4,\mu,t\right),
                                         \label{R 32}
\end{equation}
 where the
guiding center distribution functions $f_{g\nu}$ 
are constants of motion and solutions of the drift
kinetic differential equations 
\begin{equation}
 \frac{\partial f_{g\nu}}{\partial t}+{\bf v}_{g\nu}\cdot
 \frac{\partial f_{g\nu}}{\partial{\bf x}}
        +V^4\frac{\partial f_{g\nu}}{\partial q^4}= 0.
\end{equation}

In the present paper,  the second-order perturbation energy 
is calculated for the case of the equilibria defined in Sec. II
and for initial perturbations
${\bf A}^{(1)}=\dot{\bf A}^{(1)}={\bf 0}$.
It is also shown   {\sl a posteriori} that  one
can choose initial perturbations without changing the
particle contribution to the energy in a way such that
 the corresponding charge density $\rho^{(1)}$ vanishes.
Therefore, choosing initial perturbations of this kind,
  we can put from the outset
 \begin{equation}
 F_{\mu\lambda}^{(1)}\equiv 0,\ A_{\rho}^{(1)}\equiv 0.
 \end{equation}
 Equation (\ref{P2}) then reduces to
\begin{equation} 
\frac{\partial S^{(1)}_{\nu}}{\partial t} = - \left[S^{(1)}_{\nu}
H^{{(0)} }_{\nu}\right],
                                                    \label{60a}
\end{equation} 
 and the Dirac Hamiltonians to
\begin{equation} 
H^{{(0)} }_{\nu} = e_{\nu} \phi^{\star{(0)} }_{\nu} +
{\bf v}^{{(0)} }_{g\nu}\cdot\left[{\bf P} - \frac{e_{\nu}}{c}
{{\bf A}}^{\star{(0)} }_{\nu}\right].
                                               \label{61}
\end{equation} 
The second-order perturbation energy $F^{(2)}$  (Eq. (\ref{P0}))  takes then
the form
\begin{eqnarray}
F^{(2)}& =& - \sum_\nu\int d^3xdq^4d\tilde{P} 
\frac{S^{(1)}_{\nu}}{\partial t}
\left(f^{{(0)}}_{\nu} 
\frac{\partial S^{(1)}}{\partial\tilde{P}_i}\right)_{ ,i} \nonumber \\
 & &+ \sum_{\nu}
\int d^3xdq^4d\tilde{P} f^{{(0)}}_{\nu} \left({\cal H}^{(2)}_{\nu} -
{\cal H}^{(0)(2)}_{\nu}\right),
                                           \label{str}
\end{eqnarray}
with
\begin{equation}
\left(f^{{(0)}}_{\nu}
\frac{\partial S^{(1)}_{\nu}}{\partial\tilde{P}_i}\right)_{\ ,i}
= \frac{\partial}{\partial\tilde{q}^i} \left(f^{{(0)}}_{\nu}
\frac{\partial S^{(1)}_{\nu}}{\partial\tilde{P}_i}\right) 
+\frac{1}{q^{1}} f^{{(0)}}_{\nu}
\frac{\partial S^{(1)}_{\nu}}{\partial P_1},
                                          \label{59}
\end{equation}
${\bf x}(q^1,q^2,q^3)={\bf x}(r,\theta,z)$ and 
$d^3x=q^1dq^1dq^2dq^3=r drd{\theta}dz$.
After a lengthy derivation, which is presented in Appendix A,
 Eq. (\ref{str})  can be cast in the concise form
\begin{eqnarray} 
F^{(2)}& = &- \sum_{\nu}\int S(r)dr dv_\parallel d{\mu}
\left\{\frac{B^{\star{(0)} }_{\nu}}{m_\nu} \left|G^{(1)}_{\nu}\right|^2
\left({\bf k}_{\theta z}\cdot{\bf v}^{{(0)} }_{g\nu}\right) \right. 
                                                      \nonumber \\ 
& &\left.\times \left[ \left(k_{\parallel}+ k_\perp
\frac{v_\parallel}{\omega^{\star{(0)} }_{\nu}}
\frac{\left(b^{(0)}_\theta\right)^2}{r}\right)
\frac{\partial f^{{(0)} }_{g\nu}}{\partial v_\parallel} 
 - k_\perp
\frac{1}{\omega^{\star{(0)} }_\nu}
\frac{\partial f^{{(0)} }_{g\nu}}{\partial r}\right]\right\}.
                                                    \label{85}
\end{eqnarray} 
Here, $S(r)$ is a normalization surface (Eq. (\ref{76})),
$G^{(1)}_{\nu}(r,q^4,\mu)$ are arbitrary first-order functions related to
the perturbations (Eq. (\ref{g73})); 
${\bf k}_{\theta z}$, $k_\parallel$ and $k_\perp$ are the wave
vector lying in magnetic surfaces (Eq. (\ref{74})) and its components parallel
and perpendicular to ${\bf B}^{(0)}$.
We note that $F^{(2)}$
depends on $G^{(1)}_{\nu}$ only via $|G^{(1)}_{\nu} |^2$.
 
Since the first-order charge density $\rho^{(1)}$ is a $v_\parallel$ and 
${\mu}$ integral over an expression that is linear in $S^{(1)}_{\nu}$
and therefore also linear in $G^{(1)}_{\nu}$, one can satisfy
the relation ${\rho}^{(1)}=0$ 
 by a proper distribution of positive
and negative values of $G^{(1)}_{\nu}$, on which $F^{(2)}$ does not
depend. 

For a vanishing field line curvature $(B_{\theta}^{(0)} =0$ or
 $r\rightarrow\infty$),
Eq. (\ref{85}) reduces to the $F^{(2)}$ expression for plane
equilibria which was derived previously
\cite{ThPf} (Eq. (82) therein).
New terms here are the curvature-drift component of 
${\bf v}^{{(0)} }_{g\nu}$, and 
 $k_\perp\frac{\textstyle   v_\parallel}
              {\textstyle   \omega^{\star{(0)} }_{\nu}}
\frac{\textstyle  \left(b_\theta^{(0)}\right)^2}{\textstyle   r}
\frac{\textstyle   \partial f^{{(0)} }_{g\nu}}{\textstyle   
                 \partial v_\parallel}.$
 The latter term signifies that
$\frac{\textstyle  \partial f^{{(0)} }_{g\nu}}
                 {\textstyle  \partial v_\parallel}$ 
plays a role for perturbations propagating not parallel to ${\bf
B}^{{(0)}  }(k_\perp\neq 0)$, a property arising
from the fact that the curvature
drift component of ${\bf v}^{{(0)} }_{g\nu}$ depends (quadratically) on
the parallel velocity $v_\parallel$.
\begin{center}  
{\bf IV.\ \ CONDITIONS FOR THE EXISTENCE\\
 OF NEGATIVE-ENERGY PERTURBATIONS}
\end{center}
  
 First it is again noted that the conditions for the existence of
negative-energy perturbations hold if the chosen frame of
reference is that of minimum energy.  Perturbations
propagating parallel, obliquely and perpendicularly to ${\bf
B}^{(0)} $ are separately considered.
\begin{center}
{\bf A.\ \ Parallel modes $(k_\perp=0)$}
\end{center}
In this case Eq. (\ref{85}) reduces to
\begin{eqnarray} 
F^{(2)} &=&-S\sum_\nu\int rdr d v_\parallel d\mu
\left[\frac{B^{\star{(0)} }_{\nu\parallel}}{m_\nu} 
\left|G^{(1)}_\nu\right|^2
k^2_\parallel\right.     \nonumber \\ 
 &&\times
\left.  v_\parallel  
\frac{\partial f^{{(0)} }_{g\nu}}{\partial  v_\parallel}\right].
                                              \label{87}
 \end{eqnarray} 
Thus,  one obtains $F^{(2)}<0$ if
\begin{equation} 
  v_\parallel  
\frac{\partial f^{{(0)} }_{g\nu}}{\partial  v_\parallel}>0
                                                  \label{87a}
\end{equation}  
holds for some $r$, $ v_\parallel$ and
$\mu$ for any particle species $\nu$.  Condition (\ref{87a}), 
first derived by Pfirsch and Morrison \cite{PfMo} for a
homogeneous, magnetized plasma, guarantees the existence of
negative-energy perturbations without any restrictions on the
magnitude or  orientation of the wave vector other than
$k_\parallel\neq0$: it suffices to localize $G^{(1)}_\nu$ to the
region in $r$, $ v_\parallel$ and $\mu$ where
$v_\parallel
\frac{\textstyle  \partial f^{(0)} _{g\nu}}
        {\textstyle  \partial  v_\parallel}>0$. 
 Outside this region
$G^{(1)}_\nu$ vanishes.  All the other $G^{(1)}_ \lambda$,
i.e. with $\lambda\neq\nu$, are set equal to zero.  The sign
of $F^{(2)}$ is then determined only by the sign of the
integrand in the region of localization.  
 This
result agrees with those obtained by Correa-Restrepo and Pfirsch for 
several Vlasov-Maxwell equilibria \cite{CoPf92}-\cite{CoPf94b}.
\begin{center}
{\bf B.\ \ Oblique modes $(k_\parallel\neq 0\ \mbox{and}\ \ k_\perp\neq0)$}
\end{center}
With the definitions 
\begin{equation} 
C= v_\parallel\frac{k_\parallel}{k_\perp} 
 -
\frac{\mu c}{e_ \nu B^{\star{(0)} }_{\nu\parallel}} 
\frac{dB^ {(0)} }{dr}
+
\frac{v_\parallel}{\omega^{\star{(0)} }_ \nu }
\frac{\left(b^{(0)}_\theta\right)^2}{r}
                                                  \label{88}
\end{equation} 
and
\begin{equation} 
D=\frac{k_\parallel}{k_\perp} 
\frac{\partial f_{g\nu}}{\partial  v_\parallel} -
\frac{1}{\omega^{\star{(0)} }_ \nu} \left[
\frac{\partial f^{{(0)} }_{g\nu}}{\partial r} - 
v_\parallel\frac{\left(b^{(0)}_\theta\right)^2}{r}
\frac{\partial f^{{(0)} }_{g\nu}}{\partial  v_\parallel}\right],
                                                 \label{89}
\end{equation} 
Eq. (\ref{85}) yields $F^{(2)}<0$ if
\begin{equation} 
C>0 \ \ \mbox{and}\ \ D>0
\label{90a}
\end{equation} 
or
\begin{equation} 
C<0 \ \ \mbox{and}\ \ D<0.
                                    \label{90b}
\end{equation} 
The following two cases are now considered:  \newline\newline
a)\ \   Let us first assume that
\begin{equation} 
  v_\parallel  
\frac{\partial f^{(0)} _{g\nu}}{\partial  v_\parallel}>0
                                     \label{91}
\end{equation} 
again holds locally in $r$, $ v_\parallel$ and $\mu$ for any particle
species $\nu$. It then follows from inequalities (\ref{90a}) 
and (\ref{90b}) that 
\begin{equation} 
\frac{k_\parallel}{k_\perp}<\min(\Lambda_\nu,\ M_\nu)\ \ \ \mbox{or}\ \
\ \frac{k_\parallel}{k_\perp}>\max(\Lambda_\nu,\  M_\nu),
                                             \label{95b} 
\end{equation} 
with
\begin{displaymath} 
\Lambda_{\nu} \equiv \frac{1}{v_\parallel}
\frac{\mu c}{e_\nu B^{\star{(0)} }_{\nu\parallel}} 
\frac{dB^{{(0)} }}{dr} -
\frac{1}{\omega^{\star{(0)} }_{\nu}} 
v_\parallel
\frac{\left(b^{(0)}_\theta\right)^2}{r}
\end{displaymath} 
and
\begin{displaymath} 
M_{\nu} \equiv -
\frac{v_\parallel}{\omega^{\star{(0)} }_{\nu }}
\frac{\left(b^{(0)}_\theta\right)^2}{r}
+ \frac{1}{\omega^{\star{(0)} }_{\nu}}
\frac{\partial f^{(0)} _{g\nu}}{\partial r}
\left(\frac{\partial f^{(0)} _{g\nu}}{\partial q^4}\right)^{-1}.
\end{displaymath} 
The perturbations $G^{(1)}_\nu$ are localized
as in the previous case of parallel
propagation. The orders  of magnitude of
$\Lambda_\nu$ and $M_\nu$ depend on the
particle energy.  For thermal particles, these
being the most representative particles, it
holds that 
$$ |\Lambda_\nu| \approx |M_{\nu}|
\approx \frac{\textstyle   (r_{L\nu})_{th}}{\textstyle  
r_0} <<1 $$
($\frac{\textstyle   (r_{L\nu})_{th}}{\textstyle   r_0}$ 
is the thermal Larmor radius),
and consequently condition (\ref{95b}) imposes no essential restriction
on the magnitude or orientation of ${\bf k}_{\theta z}$ associated with
 negative-energy perturbations.\newline\newline
b)\ \  On the other hand, if
\begin{equation} 
  v_\parallel   
\frac{\partial f^{{(0)} }_{g\nu}}{\partial  v_\parallel} <0,
                                              \label{97a}
\end{equation}
holds 
at some $r$, $ v_\parallel$ and $\mu$ for any $\nu$, a condition which
is more frequently satisfied (e.g. in the case of  Maxwellian
distribution functions), it follows from inequalities (\ref{90a}) and 
(\ref{90b}) that
negative-energy perturbations exist if, in addition to (\ref{97a}),
\begin{equation} 
\min(\Lambda_\nu,M_\nu) <\frac{k_\parallel}{k_\perp}<\max(\Lambda_\nu,\
M_\nu)
                                                  \label{98}
\end{equation} 
holds. For thermal particles  the latter
condition  implies that
\begin{equation} 
\frac{k_\parallel}{k_\perp} \approx \frac{(r_{L\nu})_{th}}{r_0} <<1.
 \label{99}
\end{equation} 
Therefore, the most important negative-energy perturbations,
in the sense that the less restrictive condition (\ref{97a}) is
involved, concern nearly perpendicular modes.
\begin{center}
{\bf C.\ \  Perpendicular modes $(k_\parallel=0)$}
\end{center}
In this case, with the aid of the equilibrium condition (\ref{51}),
Eq. (\ref{85}) reduces to
\begin{equation} 
F^{(2)}=4\pi S\sum_\nu\int rdrd v_\parallel d\mu|G^{(1)}_\nu|^2
\frac{B^{\star{(0)} }_{\nu\parallel}}{m^2_\nu} 
\frac{W_{\nu\perp}}{\left(B^{(0)} \right)^2}
\left(\frac{k_\perp}{\omega^{\star{(0)} }_\nu}\right)^2 R_\nu
Q_\nu
                                          \label{100}
\end{equation} 
with
\begin{equation} 
R_\nu=\frac{dP^{(0)} }{dr} + \frac{(B^{{(0)} }_{\theta})^2}{4\pi r}
\left(1+ \frac{2W_{\nu\parallel}}{W_{\nu\perp}}\right) + \Pi(r)
                                            \label{101}
\end{equation} 
and
\begin{equation} 
Q_\nu= \left( 
\frac{\partial f^{{(0)} }_{g\nu}}{\partial r} -
\frac{(b^{{(0)} }_{\theta})^2}{r} \frac{\partial
f^{{(0)} }_{g\nu}}{\partial  v_\parallel}\right).
                                            \label{102}
\end{equation} 
Here, $W_{\nu\parallel}$ and $W_{\nu\perp}$ are the parallel and perpendicular
particle energies.
Negative-energy perturbations exist whenever either of the conditions
\begin{equation} 
R_\nu<0\ \ \mbox{and}\ \ Q_\nu >0
                                            \label{104}
\end{equation} 
or 
\begin{equation} 
R_\nu>0\ \ \mbox{and}\ \ Q_\nu <0
                                             \label{105}
\end{equation} 
hold.
Condition (\ref{105}), which cannot be satisfied by plane equilibria
with singly peaked pressure profiles for which
$R_\nu=\frac{\textstyle   dP^{(0)} }{\textstyle   dr}\leq 0$,
 determines a new regime of
negative-energy perturbations.  The consequences of 
(\ref{104}) and (\ref{105}) for straight tokamak and reversed-field pinch
equilibria are examined in Sec. V.  To simplify the notation, the superscript
(0) will be suppressed on the understanding that all quantities pertain to
equilibrium.
\begin{center}
 {\bf V.\ \ PERPENDICULAR NEGATIVE-ENERGY  PERTURBATIONS
 \\IN EQUILIBRIA OF MAGNETIC CONFINEMENT SYSTEMS}
\end{center} 
\begin{center}
{\bf A.\ \ Straight tokamak equilibria}
\end{center}

Straight tokamak plasmas which are close to thermal equilibrium
can be described by shifted Maxwellian
distribution functions 
\begin{equation} 
f_{g\nu}=\left(\frac{m_\nu}{2\pi}\right)^\frac{1}{2}
\frac{N_{\nu}(r)}{T_{\nu}^{3/2}(r)}    
  \exp\left\{- \frac{\mu B(r)+\frac{1}{2m_\nu}
\left[v_\parallel-V_{\nu}(r)\right]^2}{T_\nu(r)}\right\},
                                  \label{106}
\end{equation} 
where $N_\nu$ and $T_\nu$ are, respectively, the number density and
temperature (in energy units) for particles of species $\nu$.
The shift velocity $V_\nu$  satisfies
\begin{equation} 
\frac{V_\nu}{(v_\nu)_{th}} \approx
\frac{(r_{L\nu})_{th}}{r_0}\ll 1
                                    \label{107}
\end{equation} 
and, as shown later, leads to a net ``toroidal" current.

In the remainder of the paper the analysis will be carried out
up to zeroth order in $(r_{L\nu})_{th}/r_0$, i.e. small terms
of the order of
$\left[\frac{\textstyle  (r_{L\nu})_{th}}{\textstyle   r}\right]^n$ (with
$n\geq 1$) will be dropped.  In this context,
 from
Eq. (\ref{53})
one   obtains
$\Pi_\nu\approx 0$, and Eqs. (\ref{51}) and
(\ref{101}) reduce,  respectively,  to
\begin{equation} 
\frac{d}{dr} \left(P+\frac{B^2}{8\pi}\right) +
\frac{B^2_\theta}{4\pi r}=0
\label{109a1}
\end{equation} 
and
\begin{equation} 
R_\nu=\frac{dP}{dr} + \frac{B^2_\theta}{4\pi r}
\left(1+2\frac{W_{\nu\parallel}}{W_{\nu\perp}}\right).
                                   \label{109b1}
\end{equation} 
For distribution  functions (\ref{106}), negative-energy
perturbations exist if the relation
\begin{equation} 
R_\nu Q_\nu= R_\nu
\left(\frac{N^\prime_\nu}{N_\nu}\right)U_\nu f_{g\nu}<0
                                     \label{110a1}
\end{equation} 
is satisfied.
Here,
\begin{equation} 
U_\nu\equiv 1-\frac{3}{2} \eta_\nu +\eta_\nu
\frac{W_{\nu\perp}}{T_\nu}
\left(1+\frac{W_{\nu\parallel}}{W_{\nu\perp}}\right) +
\frac{4\pi}{B^2} \frac{W_{\nu\perp}}{T_\nu}
\frac{N_\nu}{N^\prime_\nu} R_\nu,
                                        \label{111a1}
\end{equation} 
with
\begin{equation} 
\eta_\nu\equiv \frac{\partial \ln T_\nu}{\partial \ln N_\nu}.
\end{equation} 
It is now assumed that both the density and
temperature profiles are singly peaked and therefore $\eta_{\nu}\geq 0$ 
for all
$\nu$. Negative-energy perturbations thus exist in the following two
regimes:\newline\newline
 a)\ \ \underline{$R_\nu<0$}.\ \  
This  implies that
$R_\nu\left(\frac{\textstyle   N^\prime_\nu}{\textstyle   N_\nu}\right)>0$ and,
consequently, condition (\ref{110a1}) is satisfied if $U_\nu<0$. 
Since the last two terms of $U_\nu$ are non-negative and vanish for
$W_{\nu\parallel}=W_{\nu\perp}=0$, the condition 
 $U_\nu<0$
can be  satisfied if
 \begin{equation}  \eta_\nu >\frac{2}{3}
                                           \label{112}
\end{equation} 
holds for some particle species $\nu$.  The existence of
perpendicular negative-energy perturbations for any perpendicular wave
number is therefore related to the threshold value of
$2/3$ of the quantity $\eta_\nu$. As discussed in
Ref. \cite{ThPf}, this threshold value is subcritical in the
sense that it is lower than the critical value
$\eta^c_\nu\approx1$ for linear stability of 
temperature-gradient-driven modes.\newline\newline
b)\ \ \underline{$R_\nu>0$}. \ \ Condition (\ref{110a1}) is now satisfied
if  $U_\nu>0$. In this case  negative energy perturbations exist for
any $k_\perp$ without restriction on the values of $\eta_\nu$.
\newline\newline

We now find the part of the velocity space occupied by particles
associated with negative-energy perturbations (active
particles).
The particular particles with energy components
$W_{\nu\parallel}=T_\nu/2$ and
$W_{\nu\perp}=T_\nu$, and consequently with velocities equal to
the root mean square velocity
$(v_\nu)_{rms}
=\sqrt{\frac{\textstyle   3T_\nu}{\textstyle   m_\nu}} $
are first examined. For these particles, henceforth 
called {\em representative particles}, the quantity $U_\nu$  
 becomes {\em independent of $\eta_\nu$}. 
Condition $R_\nu<0,\ U_\nu<0$ is then impossible and
condition $R_\nu>0,\ U_\nu>0$ , concerning the new regime, takes the
simpler form 
\begin{equation} 
-1<\frac{4\pi}{B^2} \frac{N_\nu}{N^\prime_\nu} \left(P^\prime
+\frac{B^2_\theta}{2\pi r}\right) <0.
                                              \label{107a1}
\end{equation} 
 Condition (\ref{107a1}) guarantees  that the
representative particles are active particles.
For  particles with arbitrary velocities the part of the velocity 
space occupied by active particles is determined 
on the basis of analytic solutions 
constructed in the following way:

Inserting the distribution function
(\ref{106}) into the equilibrium equation
 (\ref{48}) and  
carrying out the integrations with respect to $v_\parallel$
and $\mu$, one obtains
\begin{equation} 
J_\theta = b_\theta \sum_\nu e_\nu N_\nu V_\nu + \frac{cb_z}{B}
P^\prime = - \frac{c}{4\pi} B^\prime_z
                                                \label{108}
\end{equation} 
and
\begin{equation} 
-J_{z} = - b_{z} \sum_\nu e_\nu N_\nu V_\nu +
\frac{cb_\theta}{B} P^\prime = - \frac{c}{4\pi} \frac{1}{r}
\left(rB_\theta\right)^\prime.
                                               \label{109}
\end{equation} 
To get some simple kind of insight,  we now restrict discussion 
to $T_i=0$, a case  often considered in the
literature, e.g. \cite{HaWa}, \cite{WaHa}.  
  For cold ions Eqs. (\ref{108}) and (\ref{109})
yield
\begin{equation} 
c\frac{b_z}{B} P^\prime -eb_\theta N_e V_e = -
\frac{{c}}{4\pi} B^\prime_z
                                              \label{110}
\end{equation} 
and
\begin{equation} 
c\frac{b_{\theta}}{B} P^\prime + eb_z N_e V_e = -\frac
{{c}}{4\pi} \frac{1}{r}
(rB_\theta)^\prime 
                                               \label{111}
\end{equation} 
with $e_e =-e$ and
\begin{equation} 
P=N_eT_e.
                                              \label{112a}
\end{equation}

Let us briefly discuss here the meaning of $V_e$: For $V_e=0$
and a constant ``toroidal" magnetic field $B_z=B_0$ one obtains
from Eq. (\ref{109}) the ``toroidal" current density
\begin{equation}
J_z =- \frac{cb_\theta}{B} P^\prime.
\label{113}
\end{equation}
On the other hand, Eq. (\ref{110}) for this case yields
$P^\prime=0$.  Hence, there is neither a pressure gradient nor
a toroidal current.
For an $r$-dependent toroidal magnetic-field component,
$B_z(r)$, and $V_e=0$, Eqs. (\ref{108}) and (\ref{109}) reduce to
\begin{equation}
-\frac{B^\prime_z}{4\pi} = \frac{b_z}{B} P^\prime,
\label{114}
\end{equation}
\begin{equation}
- \frac{1}{4\pi} (rB_\theta)^\prime =
\frac{b_\theta}{B} P^\prime.
\label{115}
\end{equation}
For $B_\theta \neq 0$, one can readily show that their
solutions satisfy the relation $B_\theta  =\frac{\textstyle c
B_{z}}{\textstyle r}$ (with $c$=const.) and therefore they are singular at
$r=0$.  For $B_\theta \equiv 0$, Eq. (\ref{115}) is trivially
satisfied and Eq. (\ref{114}) describes a shearless stellaratorlike
configuration with vanishing toroidal current, a case which was
studied in Ref. \cite{ThPf}. 

To obtain analytic straight tokamak equilibria, 
 it is convenient to use, instead of
Eqs. (\ref{110}) and (\ref{111}), Eq. (\ref{110}) and
\begin{equation} 
\nabla^2 \psi =-4\pi \frac{d}{d\psi} \left(P(\psi)
+\frac{B^2_z(\psi)}{8\pi}\right),
                                      \label{116}
\end{equation} 
which is equivalent to the equilibrium condition (\ref{109a1}). 
Here, $\psi(r)$  is the usual poloidal flux function.
Assigning
the $\psi$-dependence of the functionals $P(\psi)$ and
$B_z(\psi)$ and the $r$-dependence of $V_e(r)$, one 
 obtains from the solution of Eq. (\ref{116}) the
poloidal magnetic field 
${\bf B}_\theta ={\bf \nabla}\psi \times{\bf e}_z = -
\frac{\textstyle   d\psi}{\textstyle   dr} {\bf e}_{\theta}$,
the electron density from Eq. (\ref{110})
and  the electron temperature from Eq. (\ref{112}). 
We have considered two classes of equilibria:

i)\   $B^2_z$ and $P$ are linear in $\psi$ and ii) $B_z=$ constant
 and $P=$ quadratic in
$\psi$. 
For both classes we chose  $\eta_e=0$, $\eta_e=1$,
 $\eta_e\rightarrow\infty$, and
$\eta_e<0$, the latter with singly peaked density
and  hollow temperature profiles or with singly peaked
temperature and hollow density profiles.
 From these
equilibria the following results are deduced 
(Two examples are discussed  in Appendix B.):
\begin{enumerate}
\item A substantial fraction of the thermal electrons
are active, e.g.:
\begin{itemize}
\item For linearly (marginally) stable
equlilibria of a strongly diamagnetic plasma
with $\eta_{e}=1$, more than
one-third of the thermal  electrons are active.
\item For linearly stable equilibria 
of a paramagnetic plasma with flat  electron
temperature profiles, the entire  velocity space is occupied 
by active electrons.
\end{itemize}
\item The fraction of active particles increases from
the center to the plasma edge.
\item  The fraction of active particles in a paramagnetic plasma
is higher than in a diamagnetic plasma with the same pressure
profile.
\end{enumerate}

\begin{center}
{\bf B.\ \  Reversed-field pinch equilibria}
\end{center}
 The same distribution function (\ref{106}) is
employed to derive force-free equilibria. Linearizing  Eq.
(\ref{116}) by means of the ansatz $P^\prime = 0$ and $B_z\propto \psi$ 
and then solving the resulting equation,  one obtains 
$B_{z}=B_{z}(0)J_{0}(\rho)$ and $B_{\theta}=B_{z}(0)J_{1}(\rho)$. 
These profiles   satisfactorily describe the central region of the relaxed
state of a reversed-field pinch \cite{Ta}.  We note that perpendicular
negative-energy perturbations do not exist in force-free plane equilibria
with sheared magnetic field, which were studied in Ref. \cite{ThPf},
because for this case the second-order perturbation energy vanishes.  For
cold ions and by appropriately assigning  the mean electron  velocity
profile, one can derive  equilibria  with various density and temperature
profiles having non-positive values of $\eta_{e}$ for which negative-energy
perturbations exist and a substantial fraction of active, thermal
electrons are involved.

	As an example we considered an equilibrium with  peaked density and
hollow temperature electron profiles:
\begin{equation} 
V_{e}=\mbox{const.},\ \    N_{e}=N_{e} (0) \frac{B}{B(0)},\ \     
T_{e}= T_{e} (0)
\frac{B(0)}{B}.
\end{equation} 
Condition (\ref{105}) then yields $$2\frac{W_{e\perp}}{T_{e}} +
3\frac{W_{e\parallel}} {T_{e}} <\frac{5}{2},$$ for any $\rho$, 
which implies that
more than half of the thermal electrons throughout the poloidal cross-section
are active.
\begin{center}
{\bf VI.\ \ CONCLUSIONS}
\end{center}
The general expression for the second-order perturbation energy, 
derived by Pfirsch and Morrison in the framework of linearized collisionless
Maxwell-drift kinetic theory, was
evaluated for the case of circularly cylindrical equilibria 
and vanishing initial
field perturbations. From this expression we obtained 
the following conditions for the
existence of negative-energy perturbations, 
which need only be satisfied locally in
$r$, $v_\parallel$ and $\mu$ and are valid 
in the reference frame  of minimum
equilibrium energy: 
\begin{enumerate}
  \item If the equilibrium guiding center distribution function
$f_{g\nu}^{(0)}$ of any species $\nu$ has the property  $v_\parallel 
 \frac{\textstyle  \partial f^{(0)}_{g\nu}}{\textstyle   v_\parallel } >0$,
parallel and oblique negative-energy perturbations  
$(k_{\parallel}\neq 0)$ exist with non
essential restriction on ${\bf k}$.  
\item If $v_\parallel 
\frac{\textstyle   \partial f^{(0)}_{g\nu}}{\textstyle   v_\parallel } <0$, 
the oblique
negative-energy perturbations possible are nearly perpendicular.
With the quantities
$R_\nu$ and $Q_\nu$  defined by (\ref{101}) and (\ref{102}),
the condition for perpendicular perturbations is $R_\nu Q_\nu<0$. From this
condition it follows that the curvature, 
which is associated with $B_\theta^{(0)} $,
modifies     the plane-equilibrium condition 
$\frac{\textstyle   dP^{(0)}}{\textstyle   dr}
\frac{\textstyle   \partial f^{(0)}_{g\nu}}{\textstyle  \partial r}<0$.
\end{enumerate} 
 For the case of 
tokamak equilibria there are two regimes:
\begin{enumerate}
\item If
$R_\nu<0$, the existence of negative-energy perturbations is related to the
threshold value of $2/3$ of the quantity $\eta_\nu\equiv
\frac{\textstyle  \partial \ln T_\nu} {\textstyle  \partial \ln N_\nu}$. 
\item If $R_\nu>0$, a new regime
appears, not present in plane equilibria, in which negative-energy
perturbations exist for any value of $\eta_\nu$.
\end{enumerate}
For various tokamak
 cold-ion equilibria with negative and  non-negative values of
$\eta_{e}$, a substantial fraction
of the thermal electrons are associated with negative-energy perturbations
(active particles).  In particular: 
\begin{enumerate}
\item For linearly (marginally) stable
equlilibria of a strongly diamagnetic plasma
with $\eta_{e}=1$, more than
one-third of the thermal  electrons are active.
\item For linearly stable equilibria of a paramagnetic plasma 
with flat  electron
temperature profiles, the entire  velocity space is occupied 
by active electrons.
\end{enumerate}
The part of  velocity space occupied by active particles
increases from the center to the plasma edge region and
 is larger in a paramagnetic  plasma than  in a diamagnetic
plasma with the same density and temperature  profiles. 

It is also shown that, unlike in
plane equilibria, negative-energy perturbations exist 
in force-free, reversed-field pinch
equilibria with a substantial fraction of active particles.
The present results, in particular the fact that a threshold value of
$\eta_{\nu}$ is not necessary for the existence 
of negative-energy perturbations, enhance
even more the relevance of these modes.
\begin{center} 
{\Large \bf Acknowledgments}
\end{center} 
Most of
the investigations in this  paper were conducted during a visit of
one of the authors (G.N.T.)
 to the  General Theory Division of  Max-Planck-Institut f\"ur
Plasmaphysik, Garching. The hospitality  accorded by the said institute is
gratefully acknowledged. G.N.T acknowledges 
support by the Commission of the European
Communities, Fusion Programme, Contract No. B/FUS$^\star$-913006. 
\begin{center} 
{\bf APPENDIX A:\ \ DERIVATION \\OF THE SECOND-ORDER PERTURBATION ENERGY\\
 FOR CIRCULARLY CYLINDRICAL EQUILIBRIA (Eq. (\ref{85}))
\\} 
\end{center}
\setcounter{equation}{0}
\renewcommand{\theequation}{A.\arabic{equation}}
We start from the expression  (\ref{str}).
In order that the constraints  (\ref{P4}) be satisfied,
the  terms in the first sum  of Eq. (\ref{str}), which contain
derivatives of $f^{{(0)}}_{\nu}$, are integrated by parts:
\begin{eqnarray}
\int d^3x dq^4 d\tilde{P}
\frac{\partial S^{(1)}_{\nu}}{\partial t}
\frac{\partial}{\partial \tilde{q}^i}
\left(f^{{(0)}}_{\nu}\frac{\partial S^{(1)}_{\nu}}{\partial\tilde{P}_i}\right)
&=&\int d^3x dq^4
d\tilde{P}f^{{(0)}}_{\nu}\frac{\partial S^{(1)}_{\nu}}{\partial\tilde{P}_i}\frac{\partial}{\partial q^i}\frac{\partial S^{(1)}_{\nu}}{\partial
t}                                                               \nonumber \\
& &+  \int dq^1 dq^2 dq^3d\tilde{P}f^{{(0)}}_{\nu}\frac{\partial
S^{(1)}_{\nu}}{\partial\tilde{P}_i}\frac{\partial S^{(1)}_{\nu}}{\partial t}.
                                                               \nonumber \\    
                                                               \label{62}
\end{eqnarray}
Furthermore, because of
\begin{equation}
\frac{\partial^2 {\cal H}^{{(0)}}_{\nu}}{\partial \tilde{P}_i 
\partial \tilde{P}_k} = 0,
                    \label{63}
\end{equation}
Eq. (10) of Ref. \cite{ThPf} yields
\begin{equation}
{\cal H}^{(2)}_{\nu} =0.
                                                  \label{64}
\end{equation}
Using  Eqs. (\ref{60a}), (\ref{59}),  (\ref{62}), (\ref{63}) and
Eq. (12) of Ref. \cite{ThPf} for ${\cal H}^{{(0)}(2)}_{\nu}$, and  noting that
the contribution of the
last term in (\ref{59}) cancels the contribution to  $F^{(2)}$ of 
the
last term in (\ref{62}),
Eq. (\ref{str}) is put in the form
\begin{equation}
F^{(2)} = \sum_{\nu} \int d^3x dq^4d\tilde{P}f^{{(0)}}_{\nu}
{\cal A},
                                               \label{64a}
\end{equation}
with
\begin{eqnarray}
&&{\cal A}\equiv
\frac{\partial S^{(1)}_{\nu}}{\partial\tilde{P}_i}
\frac{\partial}{\partial \tilde{q}^i}\left(
\frac{\partial H^{{(0)}}_{\nu}}{\partial\tilde{q}^i}
\frac{\partial S^{(1)}_{\nu}}{\partial \tilde{P}_j}\right) -
\frac{1}{2} (H^{{(0)}}_{\nu})_{\ ,ij}                     
\frac{\partial S^{(1)}_{\nu}}{\partial\tilde{P}_i}
\frac{\partial S^{(1)}_{\nu}}{\partial \tilde{P}_j}    \nonumber \\
&&\ \ \  - \frac{\partial S^{(1)}_{\nu}}{\partial\tilde{P}_i} 
\frac{\partial}{\partial q^i}
\left( \frac{\partial S^{(1)}_{\nu}}{\partial\tilde{q}^j}
\frac{\partial H^{{(0)}}_{\nu}}{\partial \tilde{P}_j}\right)
                                                  \label{65}
\end{eqnarray}
( $i,j=1,\ldots,4$).
To make treatment of the constraints easier, we introduce the  vector
\begin{equation} 
{\bf V}\equiv \frac{1}{m_\nu}\left[{\bf P}-\frac{e_\nu}{c}{\bf
A}^{\star{(0)} }_\nu\left({\bf x},q^4\right)\right].
                                             \label{R 34}
\end{equation} 
 It  can then be shown that
\begin{equation} 
\left.\frac{\partial H^{{(0)}}_{\nu}}{\partial\tilde{q}^j}
\right|_{{\bf V}={\bf 0}} = 0,
                                               \label{66}
\end{equation} 
\begin{equation} 
\left.\left(H^{{(0)}}_{\nu}\right)_{\ ,ij}\right|_{{\bf V}={\bf 0}} =
\left.\frac{\partial^2 H^{{(0)}}_{\nu}}{\partial q^i\partial
q^j}\right|_{{\bf V}={\bf 0}}
\label{67}
\end{equation} 
and, consequently,
\begin{equation}
\left(H^{{(0)}}_{\nu}\right)_{\ ,ij}
\frac{\partial S^{(1)}_{\nu}}{\partial\tilde{P}_i}\frac{\partial
S^{(1)}_{\nu}}{\partial\tilde{P}_j}= 
\frac{\partial S^{(1)}_{\nu}}{\partial\tilde{P}_i}
\frac{\partial}{\partial q^i}
\left(\frac{\partial H^{{(0)}}_{\nu}}{\partial\tilde{q}^j}
\frac{\partial S^{(1)}_{\nu}}{\partial \tilde{P}_j}\right).
                                                             \label{68}
\end{equation} 
We note that the constraint $P_4=0$ is not involved here, because
$P_4$ does not appear in $H^{{(0)}}_{\nu}$ (Eq. (\ref{61})).
With the aid of Eq. (\ref{68}), Eq. (\ref{65})  is now written as
\begin{equation} 
{\cal A} = \frac{1}{2} \frac{\partial
S^{(1)}_{\nu}}{\partial\tilde{P}_i}
\frac{\partial}{\partial \tilde{q}^i} \left(
\frac{\partial H^{{(0)}}_{\nu}}{\partial\tilde{q}^j} 
\frac{\partial S^{(1)}_{\nu}}{\partial \tilde{P}_j}\right)
 - \frac{\partial S^{(1)}_{\nu}}{\partial\tilde{P}_i}
\frac{\partial}{\partial q^i} 
\left( \frac{\partial S^{(1)}_{\nu}}{\partial\tilde{q}^i} \frac{\partial
H^{{(0)}}_{\nu}}{\partial \tilde{P}_j}\right).
                                                \label{69}
\end{equation} 
The two terms of  Eq.
(\ref{69}) will be calculated  separately.

The first term can be written as 
\begin{eqnarray} 
\lefteqn{\frac{\partial S^{(1)}_\nu}{\partial\tilde{P}_i}
\frac{\partial}{\partial\tilde{q}^i} \left(\frac{\partial
H^{{(0)} }_{\nu}}{\partial\tilde{q}^j}\frac{\partial
S^{(1)}_\nu}{\partial\tilde{P}_j}\right) 
=\frac{\partial^2 H^{{(0)} }_{\nu}}{\partial q^i\partial
q^j}\frac{\partial S^{(1)}_\nu}{\partial P_i} \frac{\partial
 S^{(1)}_\nu}{\partial P_j} } & &           \nonumber \\ 
& &= \frac{\partial^2
H^{{(0)} }_{\nu}}{\partial q^k\partial q^l}\frac{\partial
S^{(1)}_\nu}{\partial P_k} \frac{\partial
 S^{(1)}_\nu}{\partial P_l}  
+2\frac{\partial}{\partial q^4} \left(\frac{\partial
H^{{(0)} }_{\nu}}{\partial P^l}\right) \frac{\partial
S^{(1)}_\nu}{\partial q^l} \frac{\partial
S^{(1)}_\nu}{\partial P_4}                \nonumber \\ 
&&+\frac{\partial^2(H^{{(0)} }_{\nu}}{\partial (q^4)^2}
\left(\frac{\partial S^{(1)}_\nu}{\partial P_4}\right)^2
                                                 \label{A1}
\end{eqnarray} 
with $i,j=1,\ldots 4$ and $k,l=1,\ldots 3$.
Since the equilibrium quantities depend  just on $q^1$, the
only non-vanishing components of
$\frac{\textstyle   \partial^2H^{{(0)} }_{\nu}}
      {\textstyle   \partial q^k\partial q^l}$
and $\frac{\textstyle   \partial}{\textstyle   \partial q^4} 
\left(\frac{\textstyle  \partial
H^{{(0)} }_{\nu}}{\textstyle   \partial q^l}\right)$, according to
Hamiltonians (\ref{61}), are
\begin{equation} 
\left.\frac{\partial^2H^{{(0)} }_{\nu}}{\partial(q^1)^2}
\right|_{{\bf V}={\bf 0}}
=-\frac{e_\nu}{c} 
\frac{\partial(v^{{(0)} }_{g\nu})^l}{\partial q^1} 
\frac{\partial (A^{\star{(0)} }_{\nu})_l}{\partial q^1}
                                                  \label{A2}
\end{equation} 
and
\begin{equation} 
\left.\frac{\partial}{\partial q^4} \left(
\frac{\partial H^{{(0)} }_{\nu}}{\partial q^1}\right)
\right|_{{\bf V}={\bf 0}} 
=-\frac{e_\nu}{c}
\frac{\partial(v^{{(0)} }_{g\nu})^l}{\partial q^1} 
\frac{\partial (A^{\star{(0)} }_{\nu})_l}{\partial q^4}.
                                                  \label{A3}
\end{equation} 
 On the basis of relations (\ref{A2}), (\ref{A3}) and
\begin{equation} 
\left.\frac{\partial^2H^{{(0)} }_{\nu}}{(\partial q^4)^2}
\right|_{\bf V=0} =-\frac{e_{\nu}}{c} 
\frac{\partial(v^{{(0)} }_{g\nu})^l}{\partial
q^4} \frac{\partial (A^{\star{(0)} }_{\nu})_l}{\partial q^4},
                                                  \label{A4}
\end{equation}  
 Eq. (\ref{A1}) reduces to
\begin{eqnarray} 
\lefteqn{\frac{\partial S^{(1)}_\nu}{\partial\tilde{P}_i}
\frac{\partial}{\partial\tilde{q}^i}
\left(\frac{\partial S^{(1)}_\nu}{\partial\tilde{q}^j}
\frac{\partial H^{(0)}_\nu}{\partial\tilde{P}_j}\right)
= -\frac{e_\nu}{c} \frac{\partial(v^{{(0)} }_{g\nu})^l}{\partial q^1} 
\frac{\partial (A^{\star{(0)} }_{\nu})_l}{\partial q^1} } & &   \nonumber \\ 
&&-2\frac{e_\nu}{c} \frac{\partial(v^{{(0)} }_{g\nu})^l}{\partial q^1} 
\frac{\partial (A^{\star{(0)} }_{\nu})_l}{\partial q^4}     
-\frac{e_{\nu}}{c} \frac{\partial(v^{{(0)} }_{g\nu})^l}{\partial q^4} 
\frac{\partial (A^{\star{(0)} }_{\nu})_l}{\partial q^4}.
                                                     \label{A5}
\end{eqnarray} 
%

We now calculate the second term of Eq. (\ref{69}).
 By virtue of $\frac{\textstyle   \partial H^{{(0)} }_{\nu}}
                    {\textstyle   \partial
P_4}=0$, the second term on the right-hand side of
\begin{eqnarray} 
\frac{\partial
S^{(1)}_\nu}{\partial\tilde{P}_i}\frac{\partial}{\partial\tilde{q}^i}
\left(\frac{\partial
S^{(1)}_\nu}{\partial\tilde{q}^j}\frac{\partial
H^{{(0)} }_{\nu}}{\partial\tilde{P}_j}\right)
&=& \frac{\partial
S^{(1)}_\nu}{\partial P_i}\frac{\partial}{\partial q^i}
\left(\frac{\partial
S^{(1)}_\nu}{\partial q^l}\frac{\partial
H^{{(0)} }_{\nu}}{\partial P_l}\right)\nonumber \\ 
&&+\frac{\partial
S^{(1)}_\nu}{\partial P_i}\frac{\partial}{\partial q^i}
\left(\frac{\partial
S^{(1)}_\nu}{\partial q^4}\frac{\partial
H^{{(0)} }_{\nu}}{\partial P_4}\right)
                                             \label{A6}
\end{eqnarray} 
($i,j=1,\ldots 4,\  l=1,\ldots 3$)
vanishes.
We note here that, whereas Eq. (\ref{R 32}) for $f_\nu$ is sufficient in
the nonlinear theory to pick out the correct solutions, this is not so
with the linearized
theory. In this case, since the constraints are imposed along the
perturbed orbits, a displacement vector ({\boldmath $\xi$},$\xi_4)$
 in ${\bf x}, q^4$ space, similar to that in
macroscopic theory, is introduced \cite{PfMo};
that is, since the zeroth-order distribution function always
selects ${\bf V}={\bf 0}$ and $P_4=0$, with ${\bf V}$ 
as defined by Eq. (\ref{R
34}), it is reasonable to expand $S_\nu^{(1)}$ in powers of ${\bf  V}$ and
$P_4$:
\begin{eqnarray}
S_\nu^{(1)}&=&\hat{S}_\nu^{(1)}\left({\bf x},q^4\right)-
              {\mbox{\boldmath $\xi$}}
\cdot m_\nu{\bf  V} 
-\xi^4 P_4  \nonumber \\ 
     & &+\ \ \mbox{higher-order terms,}
                                                      \label{R 35}
\end{eqnarray}
so that
\begin{equation}
\left.\frac{\partial S_\nu^{(1)}}{\partial{\bf P}}
\right|_{{\bf  V}={\bf 0},\  P_4=0}=-{\mbox{\boldmath$ \xi$}}, 
\ \ \ \left.\frac{\partial S_\nu^{(1)}}{\partial
P_4}\right|_{{\bf  V}={\bf 0},\  P_4=0}
   =-\xi^4.
                                            \label{R 36}
\end{equation}
Using equation (\ref{R 35}), one has
\begin{eqnarray} 
\left.\frac{\partial S^{(1)}_\nu}{\partial q^l}\right|_{\bf P}
 &=&
\left.\frac{\partial S^{(1)}_\nu}{\partial q^l}\right|_{\bf V}
-\left.\frac{\partial P_k}{\partial q^l}\right|_{\bf
V}\left.\frac{\partial S^{(1)}_\nu}{\partial P_k}\right|_{\bf x}
                                               \nonumber \\ 
&=& \left.\frac{\partial S^{(1)}_\nu}{\partial q^l}\right|_{\bf V}
-\frac{e_{\nu}}{c} \frac{\partial (A^{\star{(0)} }_{\nu})_k}{\partial
q^l}\left.\frac{\partial S^{(1)}_\nu}{\partial P_k}\right|_{\bf x},
\label{A7}
\end{eqnarray} 
and, therefore,
\begin{eqnarray} 
\left.\frac{\partial S^{(1)}_\nu}{\partial q^l}\right|_{\bf P}
\frac{\partial H^{{(0)} }_\nu}{\partial P_l}
&=&\left.\frac{\partial S^{(1)}_\nu}{\partial q^l}\right|_{\bf
P}(v^{{(0)} }_{g\nu})^l                      \nonumber \\ 
&=& \left.\frac{\partial S^{(1)}_\nu}{\partial q^l}\right|_{\bf
V}(v^{{(0)} }_{g\nu})^{l}                      
-\frac{e_{\nu}}{c} (v^{{(0)} }_{g\nu})^{l}\frac{\partial
(A^{\star{(0)} })_k}{\partial q^l} \left.\frac{\partial
S^{(1)}_\nu}{\partial P_k}\right|_{\bf x}.
                                           \label{A8}
\end{eqnarray} 
Since ${\bf A}^{\star{(0)} }_{\nu}$ depends only on $q^1$ and $
{\bf v}^{{(0)} }_{g\nu}$ is perpendicular to 
${\bf e}_r$, 
 the last term on the right-hand
side of Eq. (\ref{A8}) vanishes. This has the consequence that
higher-order terms in expansion (\ref{R 35}), after
 the constraint ${\bf V}={\bf 0}$ is imposed, do not contribute to Eq.
(\ref{A9}) below. Applying the operator
$\left.\frac{\textstyle  \partial}{\textstyle  \partial q^m}\right|_{\bf P}$
($m=1,4$) to Eq. (\ref{A8}), one has
\begin{eqnarray} 
\lefteqn{\left.\frac{\partial}{\partial q^m}\left(
\frac{\partial S^{(1)}_\nu}{\partial q^l}\right|_{\bf P}
\left.\frac{\partial H^{{(0)} }_\nu}{\partial P_l}\right)\right
|_{\bf P}                                                
 =\frac{\partial}{\partial q^m}\left[
(v^{{(0)} }_{g\nu})^l\left.\left.\frac{\partial S^{(1)}_\nu}{\partial
q^l}\right|_{\bf V}\right]\right|_{\bf V}  }& &            \nonumber \\ 
&&-\frac{e_{\nu}}{c} \frac{\partial (A^{\star{(0)} }_{\nu})_k}{\partial
q^m}\frac{\partial}{\partial P_k}
\left[(v^{{(0)} }_{g\nu})^l\left.\left.\frac{\partial
S^{(1)}_\nu}{\partial q^l}\right|_{\bf V}\right]\right|_{\bf x}\nonumber \\ 
&=&\frac{\partial}{\partial q^m}\left[
(v^{{(0)} }_{g\nu})^l\left.\left.\frac{\partial S^{(1)}_\nu}{\partial
q^l}\right|_{\bf V}\right]\right|_{\bf V}\nonumber \\ 
&&-\frac{e_{\nu}}{c} 
\frac{\partial (A^{\star{(0)} }_{\nu})_k}{\partial q^m}
(v^{{(0)} }_{g\nu})^l
\frac{\partial}{\partial P_k}
\left.\left.\left(\frac{\partial
S^{(1)}_\nu}{\partial q^l}\right|_{\bf V} \right)\right|_{\bf x}
                                                \nonumber \\ 
&=& \frac{\partial}{\partial q^m}\left[
(v^{{(0)} }_{g\nu})^l\left.\left.\frac{\partial S^{(1)}_\nu}{\partial
q^l}\right|_{\bf V}\right]\right|_{\bf V}\nonumber \\ 
&&-\frac{e_{\nu}}{c} \frac{\partial (A^{\star{(0)} }_{\nu})_k}{\partial
q^m}(v^{{(0)} }_{g\nu})^l\frac{\partial}{\partial P_l}
\left(\left.\left.\frac{\partial S^{(1)}_\nu}{\partial
q^k}\right|_{\bf x}\right)\right|_{\bf V}.
                                            \label{A8a}
\end{eqnarray} 
With the  aid of expansion (\ref{R 35}) 
and (\ref{R 36}), the last
equation yields
\begin{eqnarray} 
\frac{\partial}{\partial q^m} \left[\left.\frac{\partial
S^{(1)}_\nu}{\partial q^l}\right|_{\bf P} \left.\left.\frac{\partial
H^{{(0)} }_\nu}{\partial P_l}\right]\right|_{\bf P}\right|_{\bf V=0}
 &=&
\frac{\partial}{\partial q^m}\left[
(v^{{(0)} }_{g\nu})^l\frac{\partial \hat{S}^{(1)}_\nu}{\partial
q^l}\right]            \nonumber \\  
&&+\frac{e_{\nu}}{c}\frac{\partial (A^{\star{(0)} }_{\nu})_k}{\partial
q^m} (v^{{(0)} }_{g\nu})^l \frac{\partial\xi^k}{\partial q^l}.
                                                 \label{A9}
\end{eqnarray} 
Equation (\ref{A6}) can then be written in the form
\begin{eqnarray} 
\lefteqn{\frac{\partial S^{(1)}_\nu}{\partial\tilde{P}_i}
\frac{\partial}{\partial\tilde{q}^i} \left(\frac{\partial
\hat{S}^{(1)}_\nu}{\partial\tilde{q}^j} 
\frac{\partial H^{{(0)} }_{\nu}}{\partial\tilde{P}_j}\right) 
=
\frac{\partial(v^{{(0)} }_{g\nu})^l}{\partial q^1}
\frac{\partial \hat{S}^{(1)}_{\nu}}{\partial q^l} \xi^1  } & &\nonumber \\ 
&&+(v^{{(0)} }_{g\nu})^l \frac{\partial^2 \hat{S}^{(1)}_{\nu}}{\partial
q^k \partial q^l} \xi^k
+\frac{\partial(v^{{(0)} }_{g\nu})^l}{\partial q^4} \frac{\partial
\hat{S}^{(1)}_{\nu}}{\partial q^l} \xi^4      \nonumber \\ 
 &&  +(v^{{(0)} }_{g\nu})^l
\frac{\partial^2\hat S^{(1)}_{\nu}}{\partial q^4 \partial q^l}
\xi^4
+\frac{e_{\nu}}{c} (v^{{(0)} }_{g\nu})^l
\frac{\partial(A^{\star{(0)} }_{\nu})_k}{\partial q^1}\frac{\partial
\xi^k}{\partial q^l} \xi^1                      \nonumber \\ 
&&+\frac{e_{\nu}}{c} (v^{{(0)} }_{g\nu})^l
\frac{\partial(A^{\star{(0)} }_{\nu})_k}{\partial
q^4}\frac{\partial \xi^k}{\partial q^l} \xi^4.
                                               \label{A10}
\end{eqnarray} 
Inserting of Eqs.(\ref{A5}) and (\ref{A10}) into Eq. (\ref{69}) leads to
\begin{eqnarray} 
{\cal A}&=& -\frac{1}{2} \frac{e_{\nu}}{c}
\frac{\partial\left(v^{{(0)} }_{g\nu}\right)^l}{\partial q^1}
\frac{\partial\left(A^{\star{(0)} }_{\nu}\right)_l}{\partial q^1}
\left(\xi^{1}\right)^2                                   
-\frac{e_{\nu}}{c}
\frac{\partial\left(v^{{(0)} }_{g\nu}\right)^l}{\partial q^1}
\frac{\partial\left(A^{\star{(0)} }_{g\nu}\right)_l}{\partial q^4}
\xi^1\xi^4                                 \nonumber \\ 
& & -\frac{1}{2} \frac{e_{\nu}}{c}
\frac{\partial\left(v^{{(0)} }_{g\nu}\right)^l}{\partial q^4}
\frac{\partial\left(A^{\star{(0)} }_{\nu}\right)_l}{\partial q^4}
\left(\xi^4\right)^2
    +\frac{\partial\left(v^{{(0)} }_{g\nu}\right)^l}{\partial q^1}
\frac{\partial\hat S_\nu^{(1)}}{\partial q^l} \xi^1   \nonumber \\ 
& & +\left(v^{{(0)} }_{g\nu}\right)^l 
\frac{\partial^2 \hat S^{(1)}_{\nu}}{\partial q^k \partial q^l}
\xi^k
    +\frac{\partial(v^{{(0)} }_{g\nu})^l}{\partial q^4}
\frac{\partial\hat S{(1)}_{\nu}}{\partial q^l}
\xi^4                                    \nonumber \\              
& &+(v^{{(0)} }_{g\nu})^l\frac{\partial^2 \hat
S_\nu^{(1)}}{\partial q^4 \partial q^l} \xi^4           
    +\frac{e_{\nu}}{c} \left(v^{{(0)} }_{g\nu}\right)^k
\frac{\partial\left(A^{\star{(0)} }_{\nu}\right)^l}{\partial q^1}
\frac{\partial \xi^l}{\partial q^k} \xi^1  \nonumber \\           
& & + \frac{e_{\nu}}{c} (v^{{(0)} }_{g\nu})^k
\frac{\partial(A^{\star{(0)} }_{\nu})_l}{\partial q^1}
\frac{\partial \xi^l}{\partial q^k} \xi^1                  
    +\frac{e_{\nu}}{c} \left(v^{{(0)} }_{g\nu}\right)^k
\frac{\partial\left(A^{\star{(0)} }_{\nu}\right)_l}{\partial q^4}
\frac{\partial \xi^l}{\partial q^k} \xi^4,
                                                \label{70}
\end{eqnarray} 
with $k,l=1,\ldots 3$.

 Since the equilibrium is independent on $q^2$ and $q^3$, an
appropriate ansatz for the functions $\hat S^{(1)}_{\nu}$ is
\begin{equation} 
{\hat S}^{(1)}_{\nu}\equiv G^{(1)}_{\nu} (q^1,q^4,\mu) e^{i({\bf
k}_{23}\cdot{\bf x})}.
                                               \label{g73}
\end{equation}
The wave vector ${{\bf k}}_{23} = {\bf k}_{\theta z}$ introduced here
has constant covariant components $k_2$ and $k_3$ and physical components
$k_\theta$ and $k_z$:
\begin{equation}
{{\bf k}}_{23}= k_2 \frac{\partial{\bf x}}{\partial q^2} + k_3
\frac{\partial{\bf x}}{\partial q^3} = k_{\theta} {{\bf e}}_{\theta}
+ k_z{{\bf e}}_z = {{\bf k}}_{\theta z}.
                                            \label{74}
\end{equation}
Therefore, it lies in magnetic surfaces.
We rewrite  the integral over the momentum space according
to the rule \cite{CoPfWi}
$$
\int d\tilde{P} f^{{(0)}}_{\nu} \cdots \rightarrow\int d{\mu}
 B^{\star{(0)}}_{\nu\parallel}
f^{{(0)}}_{g\nu} \cdots,
$$
 and introduce real quantities
by  \begin{equation}
AB \rightarrow \frac{1}{2} \Re A^{\star}B;
\label{75}
\end{equation}
then  inserting Eqs.  (37-40) of Ref. \cite{PfMo}
for $\xi^i$ and Eq.
(\ref{74}) into Eq. (\ref{70}), integrating with respect to $q^2$
between $q^2_0$ and $q^2_0 +\frac{\textstyle 2{\pi}}{\textstyle k_2}$
and with respect to $q^3$ between $q^3_0$ and $q^3_0+\frac{\textstyle
2{\pi}}{\textstyle k_3}$,  taking into account that
$d^3x=q^1dq^1dq^2dq^3$ 
and defining the normalization surface $S(q^1)$ by
the relation 
\begin{equation}
S(q^1)=q^1\int^{q^2_0+2{\pi}/k_2}_{q^2_0} \int^{q^3_0+2{\pi}/k_3}_{q^3_0}
dq^2dq^3,
                                 \label{76}
\end{equation}
Eq. (\ref{64a}) (after a lengthy algebra) can be written as
\begin{eqnarray}
F^{(2)} &=& \int S(q^1)d^3x dq^4d{\mu} \left\{
\frac{1}{2m_\nu} \left({\bf v}^{{(0)}}_{g\nu} 
\cdot {{\bf k}}_{23}\right) \left({\bf
B}^{\star{(0)}}_\nu\cdot {{\bf k}}_{{23}}\right) f^{{(0)}}_{g\nu}
\frac{\partial}{\partial q^4} \left|G^{(1)}_{\nu}\right|^2 \right.
                                                 \nonumber \\ 
&&-\left.\frac{c}{2e_{\nu}}
f^{{(0)}}_{g\nu}\left({\bf v}_{g\nu}^{(0)}\cdot{\bf k_{23}}\right)k_\perp
\frac{\partial}{\partial q^1} \left|G^{(1)}_{\nu}\right|^2 
       +{\cal B}\right\},
                                                            \label{77} 
\end{eqnarray}
with
\begin{eqnarray}
\lefteqn{{\cal B} \equiv \left\{ 
\frac{c}{e_\nu B^\star_{\nu\parallel}} 
\left({\bf B}^{\star{(0)}}_\nu\cdot{\bf k}_{23}\right)
\left(\frac{\partial{\bf v}^{{(0)}}_{g\nu}}{\partial q^1}
\cdot {\bf b}^{{(0)}}\right) k_\perp\right.  } & &  \nonumber \\
& &-\frac{1}{2} 
\frac{c}{e_\nu B^\star_{\nu\parallel}}
\left(\frac{\partial{\bf v}^{{(0)}}_{g\nu}}{\partial q^1} \cdot
\frac{\partial{\bf A}^{\star{(0)}}_{g\nu}}{\partial q^1}\right) k^2_\perp 
-\frac{c_\nu}{e_\nu}
\left(\frac{\partial{\bf v}^{{(0)}}_{g\nu}}{\partial q^1} 
\cdot {\bf k}_{23}\right) k_\perp              \nonumber \\
& & - \frac{1}{2m_\nu B^\star_{\nu\parallel}} 
\left({\bf B}^{\star{(0)}}_{\nu} \cdot {{\bf k}}_{23}\right)
\left(\frac{\partial{\bf v}^{{(0)}}_{g\nu}}{\partial q^4} 
\cdot{\bf b}^{{(0)}}\right)            \nonumber \\
& &\left.+ \frac{1}{m_{\nu}} 
\left({\bf B}_{\nu}^{\star{(0)}}\cdot
{\bf k}_{23}\right)
\left(\frac{\partial{\bf v}^{{(0)}}_{g\nu}}{\partial q^4} 
\cdot {{\bf k}}_{23}\right)\right\}\left|G^{(1)}_{\nu}\right|^2 
f^{{(0)}}_{g\nu} ,
                                                       \label{78}
\end{eqnarray}
\begin{equation}
k_\perp(q^1) =({\bf b}^{{(0)}} \times {{\bf k}}_{23}) \cdot
\frac{\partial {\bf x}}{\partial q^1} = ({\bf b}^{{(0)}}\times {{\bf
k}}_{\theta z}) \cdot {\bf e}_r,
                                                            \label{79}
\end{equation}
and
\begin{equation}
k_\parallel(q^1) = {\bf b}^{{(0)}}\cdot {{\bf k}}_{23}
= b^2 k_2 + b^3k_3 = b_{\theta}k_{\theta}+b_zk_z.
                                             \label{80}
\end{equation}
Integration by parts  of the terms in (\ref{77}) in which
$\partial|G^{(1)}|^2/\partial q^4$ and
$\partial|G^{(1)}|^2/\partial q^1$ appear,
leads to the expression 
\begin{eqnarray} 
F^{(2)} &=& - \sum_{\nu} \int S(q^1)d^3xdq^4d{\mu} 
\left\{\frac{1}{2m_{\nu}} \left({\bf v}^{{(0)}}_{g\nu} \cdot {{\bf
k}}_{23}\right)  \left({\bf B}_\nu^{\star{(0)}}\cdot {\bf k}_{23}\right)
|G^{(1)}|^2  \frac{\partial f^{{(0)}}_{g\nu}}{\partial q^4}
\right.                                            \nonumber \\ 
& &-\left. \frac{c}{2e_{\nu}} 
\left({\bf v}^{{(0)}}_{g\nu} \cdot {{\bf k}}_{23}\right)
k_\perp \left|G^{(1)}_{\nu}\right|^2
\frac{\partial f^{{(0)}}_{g\nu}}{\partial q^1}+{\cal B}+
{\cal C}\right\},
                                                \label{81}
\end{eqnarray}
with
\begin{eqnarray}
{\cal C}&\equiv& \left\{\frac{c}{2e_{\nu}} 
\left(\frac{\partial{\bf v}^{{(0)}}_{g\nu}}{\partial q^1} 
\cdot {{\bf k}}_{23}\right) k_\perp
+\frac{c}{2e_{\nu}}\left({\bf v}^{{(0)}}_{g\nu} 
\cdot {{\bf k}}_{23}\right) 
\left(\frac{\partial k_\perp}{\partial q^1}\right) \right.\nonumber \\
& &+\frac{c}{2e_\nu} \left(v^{{(0)}}_{g\nu} 
\cdot  {\bf k}_{23}\right)
\frac{k_\perp}{q^1} - \frac{1}{2m_{\nu}} 
\left(\frac{\partial{\bf v}^{{(0)}}_{g\nu}}{\partial q^4} 
\cdot  {\bf k}_{23}\right)(B^{\star{(0)}}_{\nu}
\cdot {\bf k}_{23})                                      \nonumber \\
& &\left.- \frac{1}{2m_{\nu}} \left({\bf v}^{{(0)}}_{g\nu} \cdot
{{\bf k}}_{23}\right) \left(\frac{\partial{\bf B}^{\star{(0)}}_{\nu}}{\partial
q^4} \cdot  {\bf k}_{23}\right)\right\}|G^{(1)}_{\nu}|^2 f^{{(0)}}_{g\nu}
.                                                   \label{82}
 \end{eqnarray}
Inserting Eqs. (\ref{40}), (\ref{44}) and (\ref{47})
 for, respectively, 
${\bf A}^{\star{(0)}}_{\nu}$, ${\bf B}^{\star{(0)}}_{\nu}$ and 
${\bf v}^{{(0)}}_{g\nu}$ into Eqs. (\ref{78}) and (\ref{82}) and
using the identities
\begin{equation}
\frac{dk_{\parallel}}{dr} \equiv -Y_{\theta z}k_\perp +
\frac{\left(b^{(0)}_\theta\right)^2}{r}
-2\frac{b_\theta^{(0)} k_\theta}{r} 
\equiv - Y_{\theta z} k_\perp 
- \frac{\left(b^{(0)}_\theta\right)^2}{r} k_{\parallel} 
+ 2\frac{b_\theta b_z}{r}k_\perp
                                                   \label{83}
\end{equation}
and
\begin{equation}
\frac{dk_\perp}{dr}  + \frac{k_\perp}{r} 
-\frac{\left(b^{(0)}_\theta\right)^2}{r} k_\perp \equiv Y_{\theta z} k_\parallel,
                                                   \label{83a}
\end{equation}
one can
show (after tedious but straightforwrd algebraic manipulations)  that
\begin{equation} {\cal B}+{\cal C}\equiv 0. 
                                                    \label{84}
\end{equation}
With the aid of Eq. (\ref{84}), Eq. (\ref{81}) reduces to Eq. (\ref{85}).
\begin{center} 
{\bf APPENDIX B: \ ACTIVE PARTICLES }
\end{center}
\setcounter{equation}{0}
\renewcommand{\theequation}{B.\arabic{equation}}
The part of the velocity space occupied by active particles 
is  determined by means of analytic solutions  of Eq. (\ref{116}).
With the ansatz $\frac{\textstyle   d}{\textstyle  
d\psi}\left(P+B_z^2\right)=\mbox{const.}$, the solution of Eq. (\ref{116})
is of the form $\psi\propto \rho^2$, with $\rho\equiv r/r_0$. This yields a
class of equilibria with the following characteristics:
\newline
 Peaked parabolic pressure profile
$$
P=P(0)(1-\rho^2),
$$
where  $\alpha$ is a parameter which describes the magnetic
properties of the plasma, i.e. the plasma is diamagnetic for
$\alpha^2<1$ and paramagnetic for $\alpha^2>1$;
\begin{equation}
B_z= \left[B^2_z(0) +8\pi P(0)(1-\alpha^2)\rho^2\right]^{1/2};
                                           \label{119}
\end{equation}
\begin{equation} 
B_\theta = 2 \sqrt{\pi P(0)} \alpha \rho;
                                          \label{120}
\end{equation} 
constant ``toroidal" current density.
Assigning appropriately the shift electron velocity profile, one can construct
equlilibria with a variety of values of $\eta_e$. Two examples are discussed
below.
\begin{center}
{\bf 1.\ \  \mbox{\boldmath$\eta_e=1$} equilibrium}
\end{center}
Choosing the $V_e$ profile as
\begin{equation}
V_e=V_e(0) \frac{B^2_f}{BB_z} (1-\rho^2)^{-1/2},
                                     \label{124}
\end{equation}
with
\begin{equation}
B^2\equiv (B^2_\theta +B^2_z) =
B^2_z(0)+4\pi P(0)(2-\alpha^2)\rho^2
                                     \label{125}
\end{equation}
and
\begin{equation}
B^{2}_{f}\equiv \left[B^2_z(0) + 4\pi
P(0)(2-\alpha^2)\rho^2\right]^{1/2},
                                     \label{126}
\end{equation}
one obtains
\begin{equation}
N_e=N_e(0)(1-\rho^2)^{1/2}\ \ \mbox{and}\ \ T_e=T_e(0)(1-\rho^2)^{1/2}.
                                     \label{127}
\end{equation}
Therefore, $\eta_e=1$ holds for all $\rho$.
We note here that, owing to the $(1-\rho^2)^{-1/2}$ dependence
of $V_e$, the equilibrium profiles are possible only in the
interval $0\leq \rho\leq \rho_s<1$, with $\rho_s$ appropriately chosen so
that inequality (\ref{107}) is satisfied (e.g.
$\rho_s=\frac{3}{4}$). Condition (\ref{107a1}),
concerning the
representative particles, yields
\begin{equation}
0<\beta(1-\rho^2)(\alpha^2-1)<1,
                                       \label{129}
\end{equation}
with
\begin{equation}
\beta \equiv \frac{P(0)}{B^2/8\pi} \approx
\frac{P(0)}{B^2(0)/8\pi} =\mbox{const.}
\label{130}
\end{equation}
The requirement that the ``toroidal" magnetic field modulus
(Eq. (\ref{125})) must be non negative sets the upper limit $1+\beta^{-1}$
on the values of $\alpha^2$ . Thus, the right
hand  inequality of condition (\ref{129})
[$\beta(1-\rho^2)(\alpha^2-1)<1$] is satisfied for all possible
values of $\alpha^2$.  The left hand inequality
[$0<\beta(1-\rho^2)(\alpha^2-1)$] is satisfied for $\alpha^2>1$ 
and, therefore, only in a paramagnetic plasma
 the representative particles are active.
For particles with arbitrary velocities, conditions
(\ref{104}) and (\ref{105}), respectively, yield
\begin{equation}
\frac{W_{e\parallel}}{W_{e\perp}} <\frac{1}{2}
\left(\frac{2}{\alpha^2}-1\right) \ \ \mbox{and} \
\left[\ 1-\frac{1}{2}
\beta(1-\rho^2)(\alpha^2-2)\right]\frac{W_{e_\perp}}{T_e} 
+\left[1-\beta\alpha^2(1-\rho^2)\right] \frac{W_{e\parallel}}{T_e}
<\frac{1}{2}
                                          \label{126a}
\end{equation}
and
\begin{equation}
\frac{W_{e\parallel}}{W_{e\perp}} >\frac{1}{2}
\left(\frac{2}{\alpha^2}-1\right) \ \
\mbox{and} \ \ 
\left[1-\frac{1}{2}
\beta(1-\rho^2)(\alpha^2-2)\right] \frac{W_{e_\perp}}{T_e} +
\left[1-\beta\alpha^2(1-\rho^2)\right]\frac{W_{e\parallel}}{T_e}
>\frac{1}{2}.
                                            \label{127a}
\end{equation}
For a strongly diamagnetic plasma
($\alpha\rightarrow0$)  condition (\ref{127a}) is
impossible and condition (\ref{126a}) yields
\begin{equation}
\left[1+\beta(1-\rho^2)\right] \frac{W_{e\perp}}{T_e} 
   + \frac{W_{e\parallel}}{T_e}<\frac{1}{2}.
                                                 \label{129a}
\end{equation}
The part of the velocity space occupied by active
particles is depicted in Fig. 1. 
 We note here that in this
and in the following figures the dotted area stands for the
active particles at plasma center $(\rho=0)$, while the area
filled by circles stands for the {\em additional} part of active
particles at $\rho=\rho_s$.
\vspace{1.5cm}
\begin{center}
\begin{picture}(7,7)
\put(0,1){\line(0,1){6}}
\put(0,1){\line(1,0){5}}
%
%
\put(3.5,1){\line(-1,1){3.5}}
\put(3.5,1){\line(-2,3){3.5}}
\put(-0.5,0.5){\large (0,0)}
\put(-1,7){\Large$\frac{W_{e\perp}}{T_e}$}
\put(5.0,0.2){\Large$\frac{W_{e\parallel}}{T_e}$}
\put(3.7,1.5){\Large$\frac{1}{2}$}
\put(0.5,6.0){\Large$\frac{1}{2a_{\small 0}(\rho_s)}$}
\put(-1.5,4.5){\Large$\frac{1}{2a_{\small 0}(0)}$}
\put(3.0,4.5){\fbox{\Large$\alpha\  \rightarrow\  0$}}
%
\multiput(.2,1.2)(.3,0){10}{\circle*{.1}}
\multiput(.2,1.5)(.3,0){9}{\circle*{.1}}
\multiput(.2,1.8)(.3,0){8}{\circle*{.1}}
\multiput(.2,2.1)(.3,0){7}{\circle*{.1}}
\multiput(.2,2.4)(.3,0){6}{\circle*{.1}}
\multiput(.2,2.7)(.3,0){5}{\circle*{.1}}
\multiput(.2,3.0)(.3,0){4}{\circle*{.1}}
\multiput(.2,3.3)(.3,0){3}{\circle*{.1}}
\multiput(.2,3.3)(.3,0){2}{\circle*{.1}}
\multiput(.2,3.6)(.3,0){2}{\circle*{.1}}
\multiput(.2,3.9)(.3,0){1}{\circle*{.1}}
%
%
\multiput(2.9,1.8)(.3,0){1}{\circle{.1}}
\multiput(2.6,2.1)(.3,0){1}{\circle{.1}}
\multiput(2.4,2.4)(.3,0){1}{\circle{.1}}
\multiput(2.2,2.7)(.3,0){1}{\circle{.1}}
\multiput(1.7,3.0)(.3,0){2}{\circle{.1}}
\multiput(1.5,3.3)(.3,0){2}{\circle{.1}}
\multiput(1.0,3.6)(.3,0){3}{\circle{.1}}
\multiput(0.8,3.9)(.3,0){3}{\circle{.1}}
\multiput(0.5,4.2)(.3,0){3}{\circle{.1}}
\multiput(0.2,4.5)(.3,0){3}{\circle{.1}}
\multiput(0.2,4.8)(.3,0){3}{\circle{.1}}
\multiput(0.2,5.1)(.3,0){2}{\circle{.1}}
\multiput(0.2,5.4)(.3,0){2}{\circle{.1}}
\multiput(0.2,5.7)(.3,0){1}{\circle{.1}}
\end{picture}
\end{center}
\vspace{.5cm}
\begin{center}
Figure 1: The part of the velocity space occupied by active electrons
 for a strongly diamagnetic plasma  with  $\eta_e=1$, wich is deduced from
Eq. (\ref{129a})\ \  [$a_0(\rho)=1+\beta(1-\rho^2)$].
 \end{center}
\vspace{.5cm}
%
Since $\beta$ is one order of
magnitude lower than unity, relation (\ref{129a}) implies that
nearly one third of the thermal electrons are active. In addition, the
fraction of active electrons slightly increases as one proceeds
from the center to the edge, because the factor $1+\beta(1-\rho^2)$ 
which multiplies $W_{e\perp}/T_e$ in relation (\ref{129a})
 is a
decreasing function of $\rho$.

For an equilibrium with constant ``toroidal" magnetic field ($\alpha^2=1$)
 conditions (\ref{126a}) and
(\ref{127a}) reduce to 
\begin{equation}
\frac{W_{e\parallel}}{W_{e\perp}} <\frac{1}{2} \ \
\mbox{and} \ \ \left[1 + \frac{\textstyle 1}{\textstyle 2}\beta(1-\rho^2)\right]
\frac{W_{e_\perp}}{T_e} +
\left[1-\beta(1-\rho^2\right]\frac{W_{e\parallel}}{T_e} <\frac{1}{2}
                                            \label{130a}
\end{equation}
and\begin{equation}
\frac{W_{e\parallel}}{W_{e\perp}} >\frac{1}{2} \ \
\mbox{and} \ \ \left[1 + \frac{\textstyle 1}{\textstyle 2} \beta(1-\rho^2)
\right] \frac{W_{e_\perp}}{T_e} +
\left[1-\beta(1-\rho^2\right] \frac{W_{e\parallel}}{T_e} >\frac{1}{2}.
                                             \label{131}
\end{equation}
The fraction of active electrons, following from conditions
(\ref{130a}) and (\ref{131}), is depicted in Fig. 2.
\vspace{1.5cm}
\begin{center}
\begin{picture}(7,7)
\put(0,1){\line(0,1){6}}
\put(0,1){\line(1,0){6}}
\put(3.5,1){\line(-1,1){3.5}}
\put(2.8,1){\line(-3,5){2.8}}
\put(0,1){\line(1,1){3.}}
\put(-0.5,0.5){\large(0,0)}
\put(-1,7.0){\Large$\frac{W_{e\perp}}{T_e}$}
\put(6.0,0){\Large$\frac{W_{e\parallel}}{T_e}$}
\put(3.5,1.5){\Large$\frac{1}{2b_1(0)}$}
\put(1.8,0.3){\Large$\frac{1}{2b_1(\rho_s)}$}
\put(.3,6){\Large$\frac{1}{2a_1(\rho_s)}$}
\put(-1.5,4.5){\Large$\frac{1}{2a_1(0)}$}
\put(2.0,4.5){$W_{e\parallel}=\frac{1}{2}W_{e\perp}$}
\put(4.2,5.8){\fbox{\Large$\alpha^2=1$}}
%
%
\multiput(.2,1.5)(.3,0){1}{\circle*{.1}}
\multiput(.2,1.7)(.3,0){2}{\circle*{.1}}
\multiput(.2,2.0)(.3,0){3}{\circle*{.1}}
\multiput(.2,2.3)(.3,0){4}{\circle*{.1}}
\multiput(.2,2.6)(.3,0){5}{\circle*{.1}}
\multiput(.2,2.9)(.3,0){5}{\circle*{.1}}
\multiput(.2,3.2)(.3,0){4}{\circle*{.1}}
\multiput(.2,3.5)(.3,0){3}{\circle*{.1}}
\multiput(.2,3.8)(.3,0){2}{\circle*{.1}}
\multiput(.2,4.1)(.3,0){1}{\circle*{.1}}
%
\multiput(3.2,1.7)(.3,0){1}{\circle*{.1}}
\multiput(2.9,2.0)(.3,0){3}{\circle*{.1}}
\multiput(2.6,2.3)(.3,0){5}{\circle*{.1}}
\multiput(2.3,2.6)(.3,0){6}{\circle*{.1}}
\multiput(2.5,2.9)(.3,0){6}{\circle*{.1}}
\multiput(2.7,3.2)(.3,0){5}{\circle*{.1}}
\multiput(3.0,3,5)(.3,0){3}{\circle*{.1}}
\multiput(3.2,3,8)(.3,0){2}{\circle*{.1}}
%
\multiput(3.0,1.3)(.3,0){1}{\circle{.1}}
\multiput(2.7,1.6)(.3,0){1}{\circle{.1}}
\multiput(2.5,1.9)(.3,0){1}{\circle{.1}}
\multiput(2.2,2.2)(.3,0){1}{\circle{.1}}
%
\multiput(1.4,3.2)(.3,0){1}{\circle{.1}}
\multiput(1.2,3.5)(.3,0){1}{\circle{.1}}
\multiput(1.0,3.8)(.3,0){1}{\circle{.1}}
\multiput(0.7,4.1)(.3,0){1}{\circle{.1}}
\multiput(0.3,4.4)(.3,0){2}{\circle{.1}}
\multiput(0.3,4.7)(.3,0){1}{\circle{.1}}
\multiput(0.1,5.0)(.3,0){1}{\circle{.1}}
\end{picture}
\end{center}
\vspace{.5cm}
\begin{center}
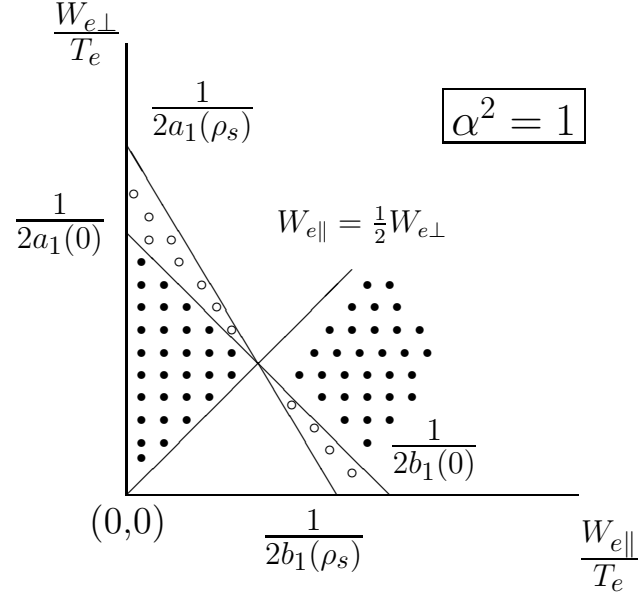
Figure 2: The part of the velocity space occupied by  active electrons
for  the equilibrium with $\eta_e=1$ and $B_z=$constant, which
is deduced from Eqs. (\ref{130a}) and (\ref{131})\ \ [$a_1(\rho) =
1+1/2\beta(1-\rho^2)$, \  $b_1(\rho) = 1-\beta(1-\rho^2$)].  \end{center}
\vspace{.5cm} 
 Nearly half of the
velocity space is now occupied by active electrons.  In addition,
one can readily show that active electrons
increase from the center to the edge.

For a diamagnetic plasma with $\alpha^2=2$   condition (\ref{126a}) is
impossible and condition (\ref{127a}) leads to
\begin{equation}
\frac{W_{e\parallel}}{T_e} 
+ \left[1 - 2\beta(1-\rho^2)\right]
\frac{W_{e_\parallel}}{T_e} >\frac{1}{2}.
                                              \label{132}
\end{equation}
The fraction of active electrons is depicted in Fig. 3.
\vspace{1.5cm}
\begin{center}
\begin{picture}(7,7)
\put(0,1){\line(0,1){5}}
\put(0,1){\line(1,0){7}}
\put(0,4.5){\line(1,-1){3.5}}
\put(0,4.5){\line(3,-2){5.2}}
\put(-0.5,0.5){\large (0,0)}
\put(-1,6){\Large$\frac{W_{e\perp}}{T_e}$}
\put(7.0,0.2){\Large$\frac{W_{e\parallel}}{T_e}$}
\put(-0.7,4.5){\Large$\frac{1}{2}$}
\put(2.5,0.2){\Large$\frac{1}{2b_{\small 2}(\rho_s)}$}
\put(4.8,2.2){\Large$\frac{1}{2b_{\small 2}(0)}$}
\put(3.0,4.5){\fbox{\Large$\alpha^2 =2$}}
%
%
\multiput(5.2,1.2)(.3,0){4}{\circle*{.1}}
\multiput(4.7,1.5)(.3,0){4}{\circle*{.1}}
\multiput(4.4,1.8)(.3,0){1}{\circle*{.1}}
\multiput(4.0,2.1)(.3,0){1}{\circle*{.1}}
\multiput(3.6,2.4)(.3,0){3}{\circle*{.1}}
\multiput(3.1,2.7)(.3,0){4}{\circle*{.1}}
\multiput(2.6,3.0)(.3,0){5}{\circle*{.1}}
\multiput(2.0,3.3)(.3,0){6}{\circle*{.1}}
\multiput(1.5,3.6)(.3,0){8}{\circle*{.1}}
\multiput(1.2,3.9)(.3,0){5}{\circle*{.1}}
\multiput(0.7,4.2)(.3,0){6}{\circle*{.1}}
\multiput(0.2,4.5)(.3,0){7}{\circle*{.1}}
\multiput(0.2,4.8)(.3,0){7}{\circle*{.1}}
\multiput(0.2,5.1)(.3,0){7}{\circle*{.1}}
\multiput(0.2,5.4)(.3,0){8}{\circle*{.1}}
%
%
\multiput(3.5,1.2)(.3,0){5}{\circle{.1}}
\multiput(3.1,1.5)(.3,0){4}{\circle{.1}}
\multiput(2.9,1.8)(.3,0){3}{\circle{.1}}
\multiput(2.6,2.1)(.3,0){3}{\circle{.1}}
\multiput(2.3,2.4)(.3,0){2}{\circle{.1}}
\multiput(2.0,2.7)(.3,0){2}{\circle{.1}}
\multiput(1.9,3.0)(.3,0){1}{\circle{.1}}
\multiput(1.5,3.3)(.3,0){1}{\circle{.1}}
\multiput(1.0,3.6)(.3,0){1}{\circle{.1}}
\end{picture}
\end{center}
\vspace{.5cm}
\begin{center}
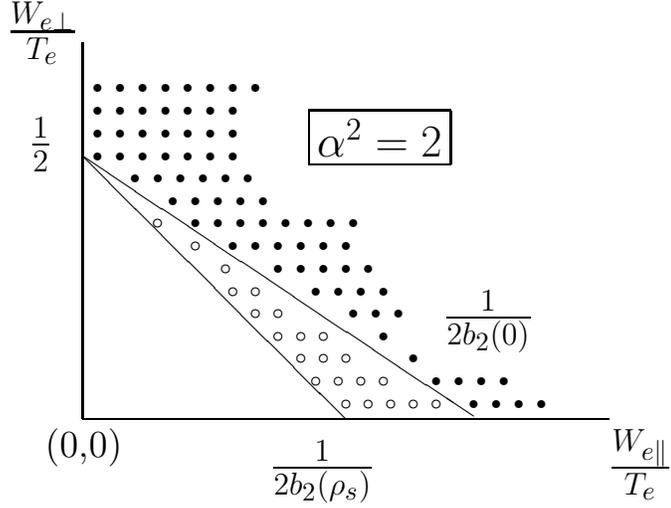
Figure 3: The part of the velocity space occupied by  active electrons
for  the equilibrium of a paramagnetic plasma  with
$\eta_e=1$, which is deduced from Eq. (\ref{132})\ \ 
[$b_2(\rho)=1-2\beta(1-\rho^2$]. 
\end{center}
\vspace{.5cm} 
Nearly
two thirds of the velocity space is now occupied 
by active electrons. 
 As in the cases of a strongly diamagnetic equlibrium and an equilibrium
with a constant `toroidal" magnetic fileld,  the fraction of active particles
increases from the center to the edge. 
\begin{center}
{\bf 2.\ \  \mbox{\boldmath$\eta_e=0$} equilibrium}
\end{center}
With the choice
\begin{equation} 
V_e=V_e(0) \frac{\textstyle   B^2_f}{\textstyle   B_z B} (1-\rho^2)^{-1}
                                                          \label{126ab}
\end{equation} 
one obtains
\begin{equation} 
N_e=N_e(0)(1-\rho^2)\ \ \ \mbox{and}\ \ \ T_e=T_e(0)=\mbox{const}.
                                                         \label{134}
\end{equation} 
Condition (\ref{107a1}) concerning the representative particles leads to
\begin{equation} 
0<\beta(\alpha^2-1)<1
                                                          \label{126b}
\end{equation} 
and, therefore, as in the equilibrium of Appendix B1, only in a paramagnetic
plasma are the representative particles active.

  For particles with
arbitrary velocities, conditions (\ref{104}) and (\ref{105})
 respectively, yield
\begin{equation} 
\frac{W_{e\parallel}}{W_{e\perp}} <\frac{1}{2}
  \left(\frac{2}{\alpha^2}-1\right)
 \ \ \mbox{and} \ \ \frac{\beta}{4}(2-\alpha^2) 
\frac{W_{e\perp}}{T_e} 
+ \frac{\beta}{2} \alpha^2\frac{W_{e\parallel}}{T_e}<-1
                                              \label{127b}
\end{equation} 
and
\begin{equation} 
\frac{W_{e\parallel}}{W_{e\perp}} >\frac{1}{2}
     \left(\frac{2}{\alpha^2}-1\right) 
\ \ \mbox{and} \ \
\frac{\beta}{4} (2-\alpha^2) \frac{W_{e\perp}}{T_e} 
+\frac{\beta}{2} \alpha^2
\frac{W_{e\parallel}}{T_e}>-1.
                                                \label{128b}
\end{equation} 
Condition (\ref{127b}) is, as expected, impossible for any $\alpha$, 
because $\eta_e$ takes its lowest non-negative value well below
the subcritical one. For $\alpha\rightarrow 0$
condition (\ref{128b}), concerning  the new
regime of negative-energy perturbations, is also impossible and
therefore
no negative-energy perturbations exist in a
strongly diamagnetic plasma. For $\alpha^2=1$ condition (\ref{128b})
yields
\begin{equation} 
\frac{W_{e\parallel}}{W_{e\perp}} >\frac{1}{2} \ \ \mbox{and} \ \
\frac{\beta}{4}  \left(\frac{W_{e\perp}}{T_e}
+ \frac{W_{e\parallel}}{T_e}\right)>-1
 \label{131b}
\end{equation} 
and, therefore, half of the velocity space is occupied by active
electrons for all $\rho$. For $\alpha^2=2$
\ condition (\ref{128b}) leads to
\begin{equation} 
\frac{W_{e\parallel}}{W_{e\perp}} >0 \ \ \mbox{and} \ \
\beta \frac{W_{e\perp}}{T_e} >-1
 \label{132b}
\end{equation} 
and therefore all particles are active.
Thus, since the value $\eta_e=0$ is far lower than the critical
value for linear stability $(n^c_e\approx1)$, negative-energy
perturbations involving a large number of thermal electrons
exist in a linearly stable regime.
\end{document}